\documentclass[journal]{vgtc}                     

\onlineid{0}

\vgtccategory{Research}

\vgtcpapertype{evaluation}

\title{Should We Dangle a Carrot? The Effect of Performance-based Incentives in Visualization Experiments}

\author{%
  \authororcid{Abhraneel Sarma}{0000-0002-1408-4511},
  \authororcid{Matthew Kay}{0000-0001-9446-0419},
  \authororcid{Sheng Long}{0009-0000-9752-5898},
  \authororcid{Michael Correll}{0000-0001-7902-3907},
  \authororcid{Alexander Lex}{0000-0001-6930-5468}
}

\authorfooter{
  \item
  	Abhraneel Sarma and Alexander Lex are with Graz University of Technology.
  \item
  	Matthew Kay and Sheng Long are with Northwestern University
  \item
  	Michael Correll is with Northeastern University
}

\abstract{%
A perennial research question in visualization involves identifying which visual encodings for a particular dataset are most effective for users in 
performing a specific task.
The relative effectiveness of the different encodings are commonly identified through controlled experiments. However, designing an experiment involves making many, often ad hoc, decisions about the experimental setup such as whether to include a training module, whether to provide performance-based incentives to participants, etc. Yet, there is limited guidance on how these decisions should be made, and we do not fully understand the impact of these subjective decisions on empirical results. In this paper, we investigate the impact of one such key design decision---monetary rewards. Specifically, we ask: does providing or not providing participants with performance-based financial incentives affect the results and the conclusions that we draw from visualization studies? We conducted two crowdsourced studies investigating the impact of incentives on (i) a low-level, perceptual task (perception of correlations in scatterplots or parallel coordinate plots), and (ii) a task involving reasoning (decision-making based on a weather forecast represented as intervals or density plots). In each of these studies, we manipulate both the visual representation and the presence of incentives as between-subject conditions. We expected to find no effect of incentives on the perceptual task, but to see an effect for the decision-making task. However, we found no effect on task performance in either study.  
While these are results of only two studies and should be replicated, they suggest that performance-based financial incentives may not always have the intended effect on participants that we presumed, and calls for a reflection of how incentivized studies should be designed.
A copy of this paper and all supplemental materials are available at \url{https://osf.io/t8eah}.
}


\graphicspath{{figs/}{figures/}{pictures/}{images/}{./}} 

\usepackage{tabu}                      
\usepackage{booktabs}                  
\usepackage{lipsum}                    
\usepackage{mwe}                       

\usepackage{mathptmx}                  

\usepackage{fontspec}
\usepackage{xspace}
\PassOptionsToPackage{dvipsnames}{xcolor}
\usepackage{lettrine}
\usepackage{scalerel}
\usepackage{xurl}
\usepackage{algorithm, algorithmic}
\usepackage{amsmath}
\usepackage{wrapfig}
\usepackage{url}
\usepackage{cellspace} 
\usepackage[nosfdefault]{sourcesanspro}
\usepackage[scaled=0.85]{FiraMono}
\usepackage{amssymb}
\usepackage[normalem]{ulem}
\usepackage{anyfontsize}

\usepackage{stfloats}

\usepackage{tikz}
\usepackage{fancyvrb}
\usepackage[T1]{fontenc}

\input{tex/helpers.tex}

\newcommand{\customsubsubsection}[1]{%
    \smallskip\noindent%
       \fontfamily{SourceSansPro-TLF}{\textbf{#1}}%
       \normalfont
}

\newcommand{\onedmatrix}[2]{\begin{bmatrix}#1 \\ #2 \end{bmatrix}}

\usepackage[hang,flushmargin]{footmisc} 

\newcounter{fncounter}
\renewcommand{\footnote}[1]{%
\footnotemark[\thefncounter]%
\footnotetext[\thefncounter]{%
#1\addtocounter{fncounter}{1}
}}

\newcommand{\supplement}{\href{https://osf.io/t8eah/overview?view_only=579c7def37de494bb08120899110af33}{\texttt{supplement}}}

\usepackage[pagebackref,bookmarks]{hyperref}
\definecolor{gray80}{HTML}{666666}

\newcommand{\ourhref}[2]{\href{#1}{%
\textcolor{gray80}{%
\texttt{#2}
\scalerel*{\includegraphics{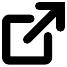}}{B}%
}%
}}

\newcommand{\preregCorr}{\href{https://aspredicted.org/p3p5py.pdf}{\texttt{preregistration}}}
\newcommand{\preregDM}{\href{https://aspredicted.org/8gm7s5.pdf}{\texttt{preregistration}}}

\newcommand{\textsans}[1]{{{\fontsize{8.65}{11}{\fontfamily{SourceSansPro-TLF}\selectfont #1}}}}

\definecolor{maroon}{HTML}{780000}
\definecolor{raspberry}{HTML}{DD2D4A}
\definecolor{seagrass}{HTML}{1a8979}
\definecolor{prussian}{HTML}{002654}

\definecolor{studyone}{HTML}{111111}
\definecolor{studytwo}{HTML}{111111}

\newcommand{\incscatter}[1]{\textcolor{maroon}{\textsans{incentivised scatter#1}}}
\newcommand{\basescatter}[1]{\textcolor{raspberry}{\textsans{base scatter#1}}}
\newcommand{\incpcp}[1]{\textcolor{seagrass}{\textsans{incentivised parallel coordinates#1}}}
\newcommand{\basepcp}[1]{\textcolor{prussian}{\textsans{base parallel coordinates#1}}}

\newcommand{\incdens}[1]{\textcolor{maroon}{\textsans{incentivised density#1}}}
\newcommand{\basedens}[1]{\textcolor{raspberry}{\textsans{base density#1}}}
\newcommand{\incci}[1]{\textcolor{prussian}{\textsans{incentivised interval#1}}}
\newcommand{\baseci}[1]{\textcolor{seagrass}{\textsans{base interval#1}}}

\newcommand{\studyone}{\textcolor{studyone}{perception of correlation}}
\newcommand{\studytwo}{\textcolor{studytwo}{decision-making}}

\definecolor{violet1}{rgb}{0.6, 0, 1}

\definecolor{gray10}{HTML}{f1f1f1}
\newcommand{\modelline}[1]{%
  \noindent\setlength{\fboxsep}{1pt}\colorbox{gray10}{\hskip0.25em\texttt{#1}\hskip0.25em}%
}

\def\hideappendix{0}

\newcommand{\appendixref}[2]{\if\hideappendix1#2\else#1\fi}

\begin{document}


\firstsection{Introduction}

\maketitle

\vspace*{-14pt}

When designing a visualization, one of the goals of a designer is to communicate the data effectively to the viewer. This typically means representing the data using visual encodings which are easy for viewers to decode. To identify the relative effectiveness of visual encodings for a particular task, visualization researchers turn to empirical studies where they compare how well participants perform on a task using two or more visual representations. The encoding which allows the average participant to perform better on a task is then considered more effective for that task. However, in the process of designing an experiment researchers need to make many, often subjective or ad hoc decisions that may have unknown effects. Here, we draw attention to the potential issue of researcher degrees of freedom in the empirical studies that visualization researchers use for determining the effectiveness of various visual encodings, and how this flexibility can impact our theoretical understanding.

This work was partly motivated by our prior experience designing an experiment to study the effect of a new visualization on a decision-making task~\cite{sarma_more_2025}. Initially, we provided participants with a description of how the data was encoded in the visualization and what the information meant; yet we observed a lot of variance in how participants interpreted the visualization and used the information to make decisions. We speculated that some of this variance could be attributable to people misunderstanding how to read the chart~\cite{nobre_reading_2024}. Subsequently, when we included more explicit information on how to read the chart during onboarding, we observed less heterogeneity in participants' responses. Since our goal was to evaluate the effectiveness of the visual representation, and not whether participants understood the representation correctly, we deployed the modified onboarding procedure.


For a typical experiment, researchers have to make several such experimental design decisions that can potentially have an impact the results. These include: how extensive the instructions should be, or whether participants should be trained on the task, given feedback, provided with performance-based monetary rewards, etc. However, these decisions are often made implicitly or even in an ad-hoc manner, and rarely articulated in a paper.
We refer to these degrees of freedom in experimental design decisions as \textit{tacit factors}. These decisions mirror the oft criticized undisclosed flexibility that researchers have in making decisions at various steps in the data analysis process~~\cite{steegen_increasing_2016, simonsohn_specification_2020},
but occur instead during the experimental design stage. 

In this work, we take a first step of investigating one such factor: \textbf{the impact of performance-based financial incentives}---rewarding correct or optimal decisions by participants through monetary bonuses---in crowdsourced visualization experiments. While some recent visualization studies, primarily on decision-making under uncertainty, have incentivized participants~\cite[e.g.,][]{kale_visual_2021, sarma_odds_2024, sarma_more_2025, fernandes_uncertainty_2018, yang_subjective_2023, padilla_uncertain_2021}, most empirical studies in visualization do not. However, \textbf{we do not know how or if the decision to incentivize participants' performance with money affects the results or the conclusions that we can draw from the studies.} This can make it tricky to compare and interpret results across studies which differingly incentivize participants but are investigating performance on the same task and using the same set of visualizations.\footnote{This tension is not merely theoretical---a real world example of this is the study by Oral et al.~\cite{oral_decoupling_2024}, which conducts a variation of the study by Kale et al.~\cite{kale_visual_2021} but without performance-based incentives. However, the two studies used different dependent variables making the results not directly comparable.} The theoretical argument for employing incentives, put forth primarily by economists, is that they induce more effort, which leads to improved performance on a task. However, we are not aware of any formal experiments on the effects of monetary incentives in crowdsourced visualization studies, a gap we attempt to close with this work.



We chose to study incentives over other \textit{tacit factors} because we expected them to be fairly straight-forward to test in a controlled experiment, and to have a significant impact on participant performance. As visualization studies encompass a wide spectrum in terms of complexity, the types of cognitive processes that are involved, etc., incentives might impact different types of studies differently. We chose two prototypical visualization studies---perception of correlation~\cite[e.g.,][]{harrison_ranking_2014, rensink_perception_2010, cutler_revisit_2026}, which involves a lower-level perceptual task; and decision-making under uncertainty~\cite[e.g.,][]{padilla_uncertain_2021, sarma_more_2025, joslyn_decisions_2013, leclerc_cry_2015}, which involves a higher-level reasoning task---to examine the effect of incentives on. We conducted two preregistered crowdsourced experiments where we varied incentives: participants were either paid a fixed amount independent of their performance, or were paid a smaller base amount, with the majority of their compensation tied to their performance.

In study 1 (\studyone), we partially replicated Harrison et al's experiment \cite{harrison_ranking_2014}. We visualized pairs of positively correlated datasets using either scatterplots or parallel-coordinates plot and asked participants to choose the plot with the greater correlation of the two shown. We expected incentives to have no impact on performance (using just noticeable difference as the metric) in this scenario, as the difference between the plots either is or is not evident, and additional effort would likely not make a difference on task performance.  

In study 2 (\studytwo), we replicated the decision-making task from Joslyn et al.~\cite{joslyn_uncertainty_2012, nadav-greenberg_uncertainty_2009, joslyn_decisions_2013}, using either an interval or a density plot to show the probability of temperatures falling below freezing, and tasked participants to make decision on whether to spend a budget to salt roads, or to preserve their budget and forgo salting. We expected incentives to have an effect on performance (using expected utility as the metric) in this scenario, as participants must make judgements about the optimal decision and estimate probabilities based on the visualized uncertainty distribution, where we expected additional effort to improve outcomes.  

In both studies, our results showed no significant performance differences between the incentivized and non-incentivized conditions. While these are results of only two studies and therefore the findings need to be replicated before stronger conclusions are drawn, they seem to suggest that, at least in the context of crowdsourced studies on platforms such as Prolific, performance-based financial incentives may have fewer benefits than we presumed. Moreover, we do find some evidence that participants spend more time performing the task in the incentivized conditions, which raises some important ethical considerations for researchers regarding ensuring fair wages for crowdworkers. We discuss the implications of our results, their possible causes, and the scope in which our results may generalize.

\section{Background}

\subsection{Researcher Degrees of Freedom in Scientific Studies}

During the course of designing, collecting and analyzing data, and reporting the results of empirical studies, researchers have to choose between several reasonable alternative choices~\cite{simmons_false-positive_2011, wicherts_degrees_2016}. In the absence of theoretically or empirically grounded guidance, they often make these choices arbitrarily. Also referred to as \textit{researcher degrees of freedom}, this flexibility, especially in the data analysis stage of a study, has received considerable attention in recent years as it can considerably increase the probability of a false-positive finding~\cite[e.g.,][]{simonsohn_specification_2020, steegen_increasing_2016, dragicevic_increasing_2019, sarma_multiverse_2023, sarma_milliways_2024, hall_survey_2022}. Our focus in this work is instead on the degrees of freedom that exist in the design stage of an empirical study (\textit{tacit factors}). These tacit factors  make it difficult to ``establish the generalizability of an effect across contexts''~\cite{landy_crowdsourcing_2020}, as, at least in the case of visualization and \textsc{hci} research, different study designs (e.g., whether or not participants are trained on how to read a chart) actually correspond to different contexts of usage~\cite{sarma_tasks_2024}.

\subsection{The Effect of Financial Incentives}\label{sec:incentive-effect}

While the effect of incentives in experiments have been explored in prior work, primarily in the field of behavioral economics, the findings have been mixed. While some studies have found incentives to improve performance, most studies have found no effect or even a detrimental effect of incentives on performance~\cite{camerer_effects_1999, cala_financial_2022, achtziger_higher_2015}. Many of the studies which found that incentives did not improve performance also found that incentives reduced variance in responses. Based on their review of prior work investigating the effect of incentives, Camerer and Hogarth~\cite{camerer_effects_1999} propose the \textit{capital-labor-production} theory to describe the potential effect of incentives in tasks. Here, \textit{capital} refers to ``cognitive capital'' or knowledge that participants possess coming into the task. \textit{labor} refers to effort that is required to perform the tasks. In most visualization and \textsc{hci} studies, participants primarily exert cognitive effort. \textit{Production} refers to the kinds of capital that is necessary for performing a task.

The \textit{capital-labor-production} theory~\cite{camerer_effects_1999} rests on a few key intuitions: (i) people dislike exerting effort, but will put in more effort if that means increased rewards; (ii) effort generally, monotonically, improves performance; (iii) capital can be a substitute for labor (for example, if one is familiar with the biases in area perception implied by Steven's power law~\cite{stevens_psychophysical_1957}, they may be more accurate at comparing areas of two circles, but someone who does not know of these biases can be just as accurate by comparing the radii and computing the areas mathematically, which entails more effort); (iv) if a task's production requirements (i.e., the effort it requires) are too low or too high, there would be little marginal gain for the participant from the extra effort that they would exert due to the presence of incentives.

One important caveat is that these studies were often conducted in physical lab spaces (not on crowdsourcing platforms), and involved very different types of tasks than what we typically employ in visualization studies. In a study conducted with crowdworkers on Mechanical Turk, Mason and Watts~\cite{mason_financial_2009} found that participants performed a greater number of tasks as pay increased, but the accuracy of the performed tasks did not improve with increases in pay. However, the tasks that participants were asked to perform in these experiments were either quite menial (ordering images temporally) or were word games, both of which are tasks where incentives are unlikely to help~\cite{camerer_effects_1999}. Other studies have investigated factors that can influence performance through an interacting effect with incentives such as intrinsic motivation~\cite{cerasoli_intrinsic_2014}. As our goal was to study the \textit{total effect of incentives}regardless of participants' intrinsic motivation, we do not control for motivation in our study design or analysis~\cite{mcelreath_statistical_2020}.

Based on the literature~\cite{camerer_effects_1999}, we may expect the effect of incentives to manifest through improved performance on certain (but not all) tasks. Specifically, the tasks which are most likely to effected by incentives are those where most or all participants possess the required cognitive capital to perform the task, and the task does not require too little or too much effort. To investigate the effect of incentives, we chose two visualization tasks---(1) perception of correlation and (2) decision-making under uncertainty---which require varying degrees of cognitive capital and effort. In addition, we may expect reduced variance in participants' responses, and an increase in the amount of time participants spend on the task (which might be a proxy for increased effort). 

\subsection{Perception of Correlation}
There has been a great deal of work in visualization and graphical perception on ranking different visual representations on their ability to help viewers accurately estimate correlation (and small differences in correlation). How accurately a viewer is able to perceive correlation using a specific representation is often determined by estimating just-noticeable differences (\textsc{jnd}s)---the minimum difference in stimuli at which a viewer can reliably (75\% of the time) detect a difference. 

Prior work has found that viewers can perceive correlation fairly accurately when the data is represented using scatterplots~\cite{rensink_perception_2010, harrison_ranking_2014}, and the \textsc{jnd}s can be described using Weber's law. Harrison et al.~\cite{harrison_ranking_2014} evaluate perceptions of correlation using other visualizations such as parallel coordinate plots, stacked bars, ordered lines etc. The work of Harrison et al.~\cite{harrison_ranking_2014}, and a re-analysis by Kay and Heer~\cite{kay_webers_2016} show that participants' perceptions of \textit{negative} correlation are almost as accurate when the data is represented using either scatterplots or parallel coordinates; however, participants' perception is much less precise for \textit{positive} correlation when the data is represented using parallel coordinates (compared to scatterplots).

Perception of correlation is often considered to be a more basic or lower-level cognitive function~\cite{harvey_domains_2019}. Prior work on perceptions of correlation have exclusively paid participants a flat amount for participating in the study. Our hypothesis was that if incentives \textit{were} to have a meaningful effect, we might observe that a visualization such as the parallel coordinates plot, which is less accurate compared to scatterplots for positive correlation in the absence of incentives, is as accurate as scatterplots when incentivized. However, based on the findings of Camerer and Hogarth~\cite{camerer_effects_1999}, we speculate that this task is unlikely to be impacted by incentives as it might not benefit from exerting additional effort. Thus, this study represents an important edge case which we seek to evaluate. In our study, we compare scatterplot and parallel coordinates plot for positive correlations. 

\subsection{Decision-making under Uncertainty}
A growing number of studies have investigated the effectiveness of different visual representations in accurately extracting probability information from uncertainty representations~\cite[e.g.,][]{kay_when_2016, hullman_hypothetical_2015, correll_error_2014}, assessing risk~\cite[e.g.,][]{ruginski_non-expert_2016, padilla_effects_2017, galesic_using_2009}, or making utility-optimal decisions~\cite[e.g.,][]{kale_visual_2021, fernandes_uncertainty_2018, sarma_odds_2024, yang_subjective_2023, joslyn_uncertainty_2012}. In this work, our focus is on the effect of visualizations in helping people make rational (i.e., decisions which are or close to utility-optimal). Prior work has found that uncertainty representations which are more expressive and present viewers with more complete distributional information such as density or violin plots, quantile dotplots, cumulative density plots lead to better decisions compared to interval plots which communicate a 95\% (or similar) confidence or credible interval~\cite{fernandes_uncertainty_2018, kale_visual_2021, sarma_odds_2024}.

In these types of studies, researchers have often used financial incentives~\cite[e.g.,][]{fernandes_uncertainty_2018, kale_visual_2021, padilla_uncertain_2021, sarma_odds_2024} based on the notion that incentivising participants will increase both the internal and ecological validity of a study~\cite{fernandes_uncertainty_2018, wu_rational_2024, kale_logic_2025}. When making decisions, people, both in the real world and in experiments, are likely 
trading off some \textit{subjective costs} against some \textit{subjective benefits}. By explicitly stating what these costs and benefits are to a participant, incentives, in theory, provide benchmarks for defining optimal or good decisions, as well as enable the evaluation of participants' decisions against the benchmark~\cite{sarma_odds_2024, wu_rational_2024}. As incentives reflect the costs and benefits associated with real-world decision-making, it also should make the findings more ecologically valid~\cite{fernandes_uncertainty_2018}, assuming that the incentives in the experiment align well with the relative costs of real-world tasks.

These types of decision-making tasks require participants to perform relatively more complex (or higher-order) reasoning tasks, which require more effort. In our study, we examine the impact of incentives by comparing how participants perform in a decision-making task where the uncertainty distribution is represented using either an interval representation---specifically 66\% and 95\% intervals---which are typically considered sub-optimal, or density plots, which are have been found to be better~\cite{fernandes_uncertainty_2018, kale_visual_2021}. Based on the \textit{capital-labor-production} theory~\cite{camerer_effects_1999}, we expect incentives to impact the results, as participants who exert more effort would be more likely to spend time and more accurately estimate the probability values from the uncertainty representation which is necessary to perform the task. Specifically, in the presence of incentives, our hypothesis was that participants will perform meaningfully better when uncertainty information is presented using density plots compared to interval plots, while in the absence of incentives, we might expect this improvement in performance to be much smaller or even disappear. 


\section{Experiment Design Preliminaries}
\label{section:experiment-prelims}
We conducted preregistered experiments on the \studyone{} and \studytwo{} tasks to study the effect of incentives. The study procedure for both studies was approved by the ethics board at Graz University of Technology (GZ EK-109/2026).

For both studies, our goal was to ensure that participants' average compensation would be approximately the same (\$15/h) across both the baseline and incentivized conditions. We first calculated the average number of correct responses for both experiments based on our pilot studies. On average, participants correctly performed the task in study 1 in ~40/65 trials. For study 2, prior work which used the same experimental setup~\cite{sarma_more_2025} showed that the average participant had 4000 virtual dollars left at the end of the study. We then assigned dollar values for answering a question correctly in study 1 (\$0.05) and for the amount of virtual dollars remaining in the bank in study 2 (\$0.5 for every \$1,000). Thus, we expected the average bonus that participants in the incentivized conditions would receive to be approximately \$2 for both studies. The maximum bonus they could theoretically receive was \$3.25 for experiments 1 and \$9 for experiment 2. We anticipated both tasks to take approximately 9-12 minutes on average, which, at the desired compensation rate of \$15/h, translates to a total compensation of \$3. Thus, the compensation for the baseline (non-incentivised) conditions was set to \$3 for both studies; the compensation for the incentivized conditions were  \$1.25 and \$1.5 for Experiments 1 and 2 respectively. The slightly higher \textit{estimated} average compensation in the incentivized conditions were set in order to comply with Prolific's minimum wage of \$8/h, which does not include bonuses.

While these amounts were the compensation advertised to participants \textit{before} taking the study, as participants took significantly longer in the incentivized conditions, we adjusted the guaranteed compensation amount to both comply with Prolific's standards and to match our desired target compensation of \$15/h. The exact compensation amounts are reported separately for each study.

\section{Experiment 1: Perception of Correlation}
We conducted a preregistered study to partially replicate the perception of correlation study by Harrison et al.~\cite{harrison_ranking_2014} using either scatterplots or parallel coordinates plot (see \preregCorr). As this task involves more lower-level cognitive functions~\cite{harvey_domains_2019}, we test this as a possible edge-case of a visualization task where incentives will likely have little to no effect on performance.

\subsection{Experimental Materials}
\begin{figure}[b!]
    \vspace{-12pt}
    \centering
    \includegraphics[width=0.9\columnwidth]{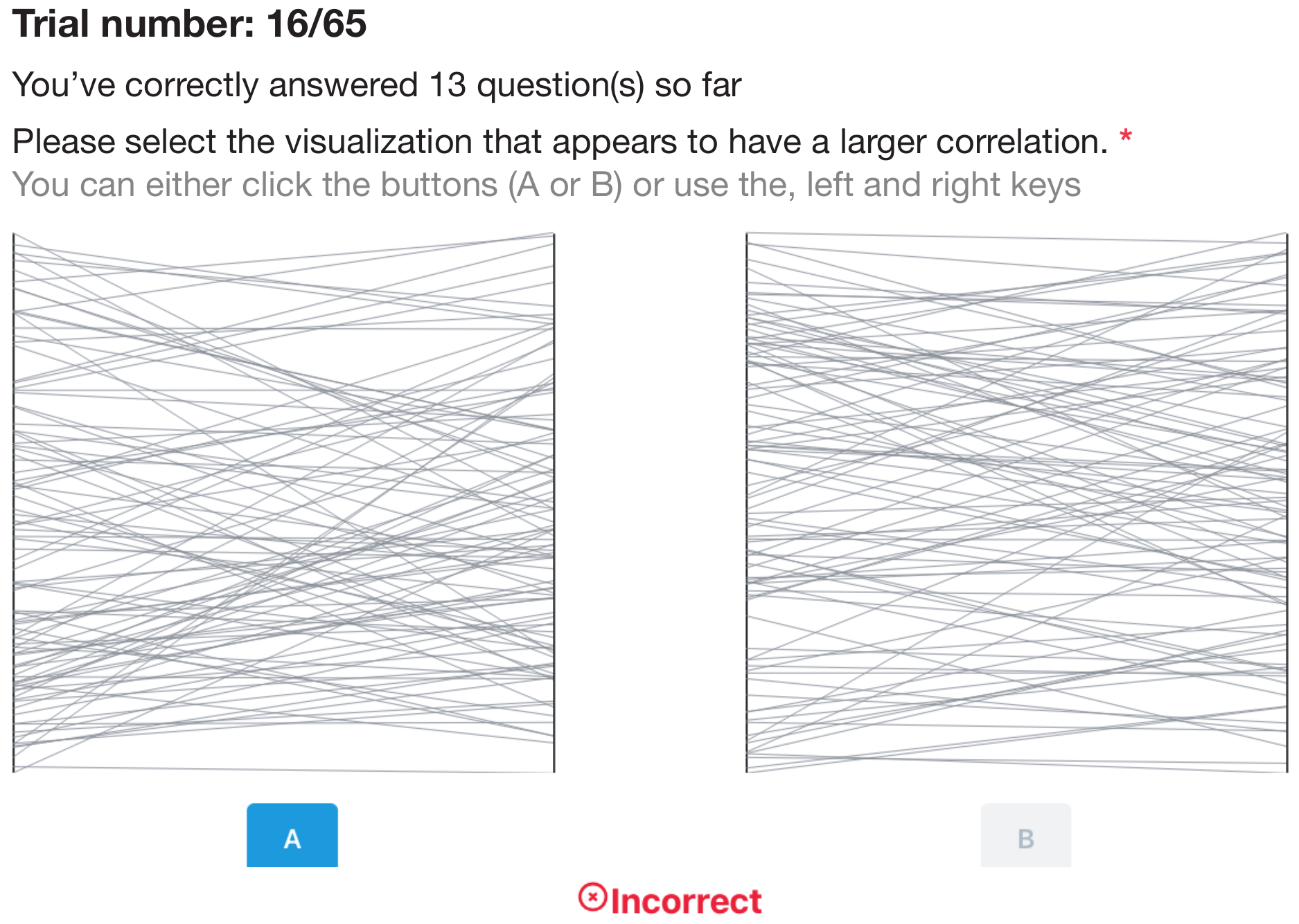}
    \vspace{-6pt}
    \caption{Example of a stimulus seen by a participant in the non-incentivized parallel coordinates plot condition.
    }
    \label{fig:pcp-example}
\end{figure}

Figure~\ref{fig:pcp-example} shows a screenshot of the experimental interface. The study is available to browse \ourhref{https://abhsarma.github.io/adversarial-effects-experiment/correlations}{here} and the source code for the study is available on \ourhref{https://github.com/abhsarma/adversarial-effects-experiment}{GitHub}. 

\customsubsubsection{Experimental Task and Conditions:} We adapted the perception of correlation task from Harrison et al.~\cite{harrison_ranking_2014} for this experiment. Like Harrison et al.~\cite{harrison_ranking_2014}, in each trial, participants were shown two charts and were asked to select the one  with the higher correlation in a Two-Alternative Forced Choice (2AFC) task. There were four experimental variables in this study: (1) the visualization (scatterplot or parallel coordinates plot); (2) the incentive scheme (advertised\footnote{For both this and the subsequent experiment, we had to adjust the guaranteed amounts in the incentivized conditions to ensure an hourly wage of approximately \$15/h, as reported in \autoref{section:experiment-prelims}. The actual amounts are reported below.} as a participation fee of \$3 or a participation fee of \$1.25 and a bonus of \$0.05 for each correct response); (3) the base correlation ($r_\textsc{base}$); and (4) the difference in correlation between the two charts ($\Delta r$). The visualization and incentive scheme were varied between-subjects whereas $r_\textsc{base}$ and $\Delta r$ were varied within-subjects. To keep the experiment simple, we only considered positive correlations, and the \textit{approach from above}---in other words the correlation of the second chart was always greater than the correlation of the first chart and is given by $r_2 = r_\textsc{base} + \Delta r$ (the order in which the charts appeared was randomized).

\customsubsubsection{Procedure:} We test five levels of $r_\textsc{base} = \{0.3, 0.4, 0.5, 0.6, 0.7\}$ and thirteen levels of $\Delta r = \{0.03, 0.04, \dots, 0.1, 0.12, 0.14, 0.18, \dots, 0.26\}$ in a fully crossed design resulting in 65 trials. In addition, we included five attention check questions where the correlations for the two charts are 0.01 and 0.99. The order of the trials were completely randomized. Unlike previous studies~\cite{rensink_perception_2010, harrison_ranking_2014, cutler_revisit_2026}, we did not employ a staircase procedure due to the possibility of a confounding effect with the incentives---we felt that there might be a possibility of two participants ending up with the same payout in the incentivized condition even if they have different \textsc{jnd}s due to the mechanisms of the staircase procedure. Unlike prior studies~\cite{harrison_ranking_2014, cutler_revisit_2026} we provided immediate feedback to participants on whether they answered correctly,
as seen in~\autoref{fig:pcp-example}. At the end of the experiment, we informed participants about their total payout. After completing all trials, participants were asked to describe the strategy that they used in performing the task, and about their prior experience with incentivized studies.

\customsubsubsection{Tutorial and Training:} Before participants began the study, we provided them with instructions on what correlations are with examples of what different levels of correlations look like, using the visualization that they will encounter subsequently. This is followed by participants completing nine training trials with feedback, using the same set of correlation stimuli that was used by Cutler et al~\cite{cutler_revisit_2026}.

\customsubsubsection{Participants:} We recruited all participants from Prolific. Our experiment was only eligible to participants who were fluent in English, and on desktop devices. We aimed to recruit 200 participants (50 participants in each condition). We excluded five participants who failed to meet our criteria, and recruited additional participants to meet our target. We received explicit consent from all participants to collect and share their responses. The median completion time for participants in the baseline condition was approximately 11.5 mins, and they were compensated \$3 (approximately \$15.75/h); the median completion time for participants in the incentivized condition was approximately 13.5 mins, and they were received a guaranteed amount of \$1.6 (adjusted up from \$1.25) and the average bonus was \$2.18 (corresponding to an average wage of \$16.8/h).

\subsection{Model specification}

We adapted the approach used by Kale et al.~\cite{kale_visual_2021}, and used a Binomial logistic regression model to estimate the probability of a correct response. The logistic fits (\modelline{line 2}) describing each participants' responses are equivalent to psychometric functions~\cite{gonzalez-rubio_webers_2026}, which can then be used to derive an estimate of \textsc{jnd} (which we discuss further below).\\

\begin{minipage}[h]{\columnwidth}
{\includegraphics{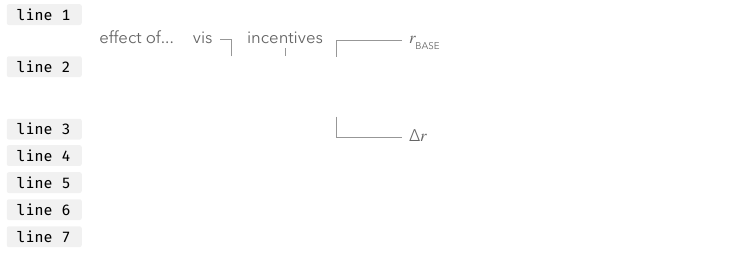}\vspace{-138pt}}
\begin{align*}
    & \hskip4em \mathit{correct} \sim \text{Binomial}(p) \\ \\
    & \hskip4em \mathrm{logit}(p) = \alpha_i + \beta_{\textsc{vis}[i]} \cdot \beta_{\textsc{inc}[i]}  \cdot \gamma_{\textsc{r}} r_{\textsc{base}[j]} + \\
    & \hskip4em \hskip6.3em \beta_{\textsc{vis}[i]} \cdot \beta_{\textsc{inc}[i]} \cdot \gamma_{\textsc{d}} \Delta r_{[k]} \\
    & \hskip4em \alpha_{i} \hskip0.5em = \hskip0.5em \alpha_0 + \delta_{\alpha, i} \\
    & \hskip4em \delta_{\alpha,i} \hskip0.5em \sim \hskip0.5em \text{Normal}(0, \sigma) \\
    & \hskip4em i \in \{1...N\} \hskip5em (N \; participants) \\
    & \hskip4em j \in \{1...J\} \hskip5em (J \; \text{levels of } r_{\textsc{base}}) \\
    & \hskip4em k \in \{1...K\} \hskip4.8em (K \; \text{levels of } \Delta = | r_{\textsc{base}} - r_2 |) \\[-8pt]
\end{align*}
\end{minipage}

\customsubsubsection{Line 1:} Whether a participant correctly chose the graph with the higher correlation in each trial is modeled as a Binomial distribution with probability $p$, where 1 means the participant made the correct choice. 

\customsubsubsection{Line 2:} For each stimuli (which consists of two graphs), we assume that participants' decisions would depend on the visualization, the incentive condition, the base correlation ($r_\textsc{base}$), and the differential increase in stimuli i.e. correlation ($\Delta r$). This model incorporates the (directionless) assumption that the probability of a correct response ($p$) will change as $r_\textsc{base}$ changes, and the magnitude of this change in $p$ will also vary with $\Delta r$. Based on prior work, both of these effects are likely to be in the positive direction.

\customsubsubsection{Lines 3-4:} We expect the intercept ($\alpha_{i}$) parameter to vary between participants, as different participants will likely have different capital and perceptive abilities. $\alpha$ is the slope parameters for the \textit{average} participant (when $\delta_{\alpha, i} = 0$); $\delta_{\alpha, i}$ captures differences between participants as random effects.

\customsubsubsection{Priors:} We use weakly-regularizing priors centered on zero (no effect) while permitting the possibility of significant distortion: $\alpha \sim \mathrm{Normal}(0, 1)$. We use zero-centered priors with a standard deviation of 2 for the random effects parameters ($\delta_{\alpha, i}$).

\customsubsubsection{Implementation:} We implemented these models in \texttt{R 4.4.0}~\cite{RLang_2024} and \texttt{CmdStanR 0.8.0}~\cite{cmdstanr}. The model ran for four chains with 5,000 warmup samples and 5,000 post-warmup samples each, thinned by 4 for a final total sample size of 5,000. We assessed convergence using the Gelman-Rubin diagnostic ($\hat{R}$ = 1.00 for all population-level parameters, correlations and standard deviations) and the (bulk and tail) effective sample sizes ($\text{ESS}_{min} \approx 2,000$).

\customsubsubsection{JND Calculation:} Based on the estimates of this model, we can calculate the JND at a specific level of $r_\textsc{base}$for a particular visualization and incentive condition as:

\begin{minipage}[h]{\columnwidth}
{\hspace{-12pt}\includegraphics{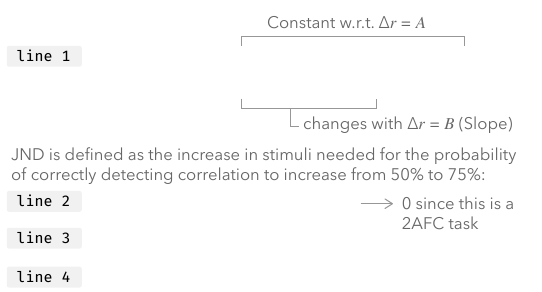}\vspace{-130pt}}
\begin{align*}
    \text{Let, } & F({\Delta r} | r_\textsc{base}) = \alpha_i + \beta_{\textsc{vis}[i]} \cdot \beta_{\textsc{inc}[i]}  \cdot \gamma_{\textsc{r}} r_{\textsc{base}[j]} + \\
    & \hskip6.2em \beta_{\textsc{vis}[i]} \cdot \beta_{\textsc{inc}[i]} \cdot \gamma_{\textsc{d}} \Delta r_{[k]} \\ \\ \\ \\
    & \mathrm{JND} = F^{-1}(0.75) - F^{-1}(0.5) \\
    & \hskip1.75em = F^{-1}(0.75) \\
    & \hskip1.75em = \frac{\mathrm{logit}(0.75) - A}{B}
\end{align*}
\end{minipage}

\subsection{Results}

\begin{figure}[b!]
    \centering
    \includegraphics[width=\columnwidth]{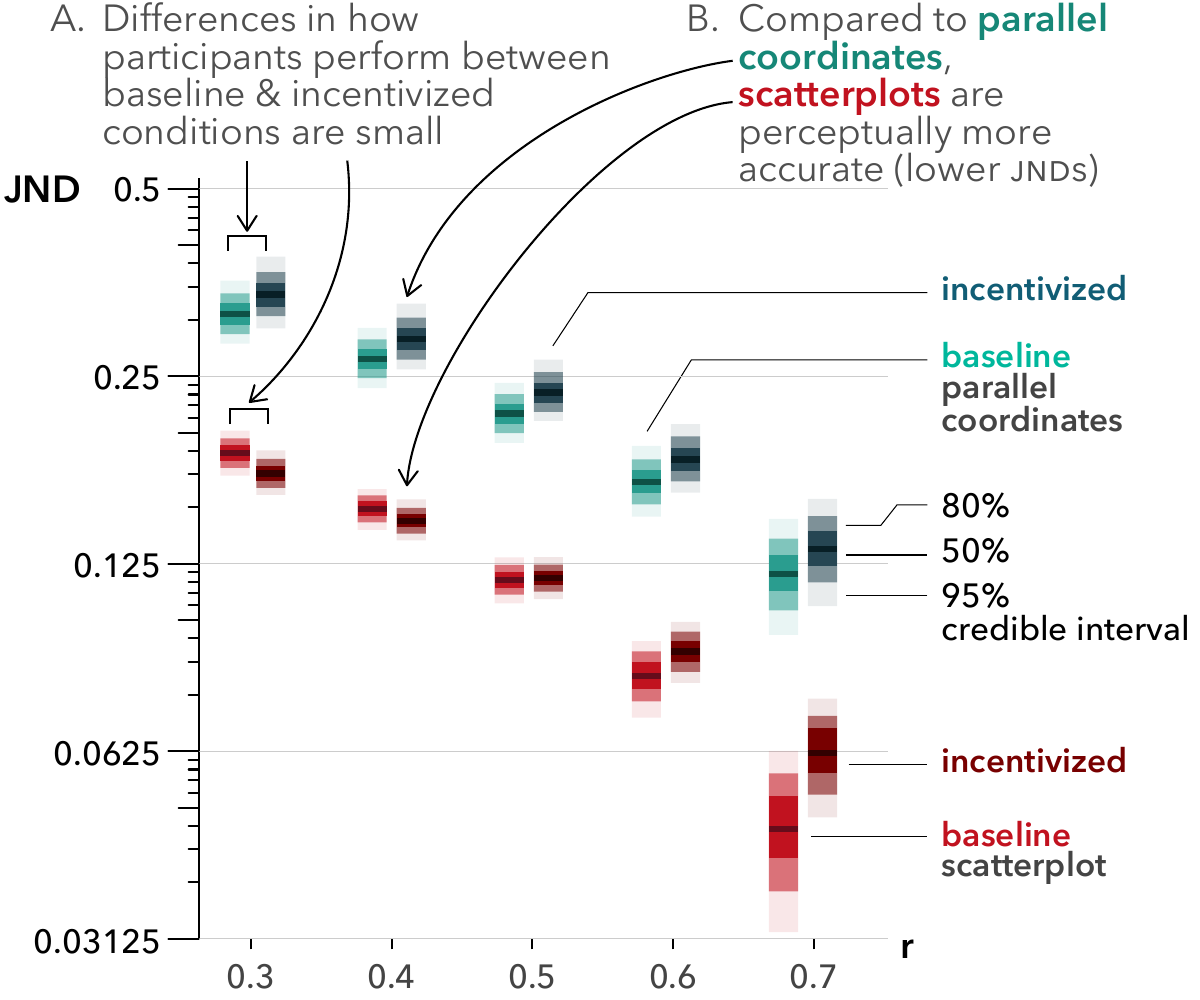}
    \vspace{-16pt}
    \caption{The main result of Experiment 1. We show the posterior estimates of \textsc{jnd} for both visual representations across the incentivized and baseline conditions, at each value of $r_\textsc{base}$.}
    \label{fig:jnd-results}
\end{figure}

The results for Experiment 1 are shown in \autoref{fig:jnd-results} as estimates of \textsc{jnd} (just noticeable differences). In the baseline (non-incentivized) condition, we find that the \textsc{jnd} for scatterplots is much lower at all levels of the baseline correlation, replicating the finding from prior work~\cite{harrison_ranking_2014, kay_webers_2016}.

\begin{figure}[t!]
    \centering
    \includegraphics[width=0.95\columnwidth]{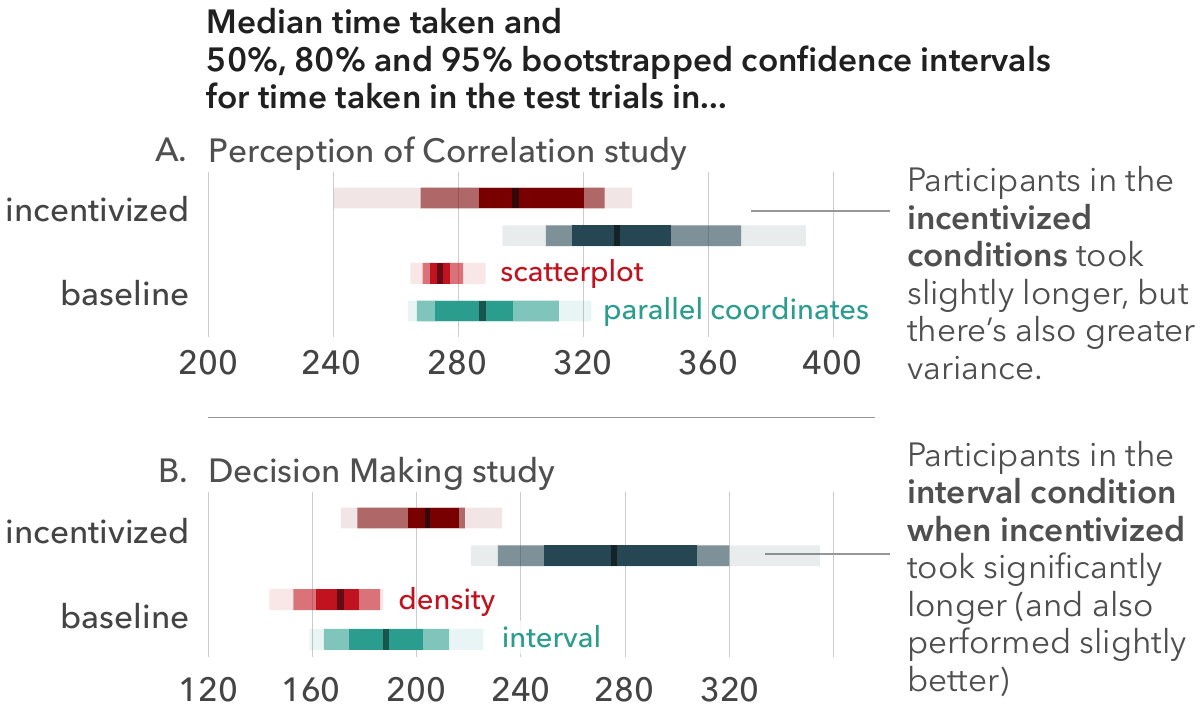}
    \vspace{-6pt}
    \caption{Median time taken and 50\%, 80\%, and 95\% bootstrapped confidence intervals for time taken in the test trials for the two studies.}
    \label{fig:time-taken}
    \vspace{-12pt}
\end{figure}

As expected, we also \textbf{do not find any evidence to suggest that incentives impact performance in the perception of correlation task}---the \textsc{jnd}s for participants in the incentivized conditions were comparable, if not slightly worse, than the \textsc{jnd}s for participants in the baseline condition (\autoref{fig:jnd-results}A). We also do not find any evidence to suggest that incentivizing participants led to reduced variance in their responses, as measured through the group-level standard deviation parameters in our multi-level logistic regression model (see \supplement{} $\blacktriangleright$ R $\blacktriangleright$ \texttt{dm-05-effect\_variance.qmd}). However, we did find that participants took slightly longer to complete the task in the incentivized conditions compared to the baseline conditions. \autoref{fig:time-taken}A, visualizes the bootstrapped median and 95\% quantile intervals for the median estimate. The difference in time spent is approximately 30s for parallel coordinates (\basepcp{: 288s, [264s, 323s]}, \incpcp{: 318s, [294s, 391s]}), and approximately 17s for scatterplot  (\basescatter{: 275s, [265s, 289s]}, \incscatter{: 292s, [240s, 336s]}). The increase, which is approximately 6-10\%, should likely be considered small.


Finally, as an aside, we want to highlight the ability of our experimental setup to replicate \textsc{jnd} estimates from prior work \textit{without using a staircase procedure}. An additional advantage of this setup is that it allows us to model participants' decisions directly using the Binomial logistic regression model described above.

\section{Experiment 2: Decision-making under Uncertainty}
\label{sec:exp2}

In this preregistered study, we examined the impact of incentives on performance in a decision-making under uncertainty task adapted from prior work~\cite{joslyn_uncertainty_2012} using either interval or density plots as uncertainty representations (see \preregDM). As performing this task involves relatively more complex cognitive functions, our hypothesis was that incentives would have an effect on performance. 

\subsection{Experimental Materials}
\begin{figure}[t!]
    \centering
    \includegraphics[width=0.95\columnwidth]{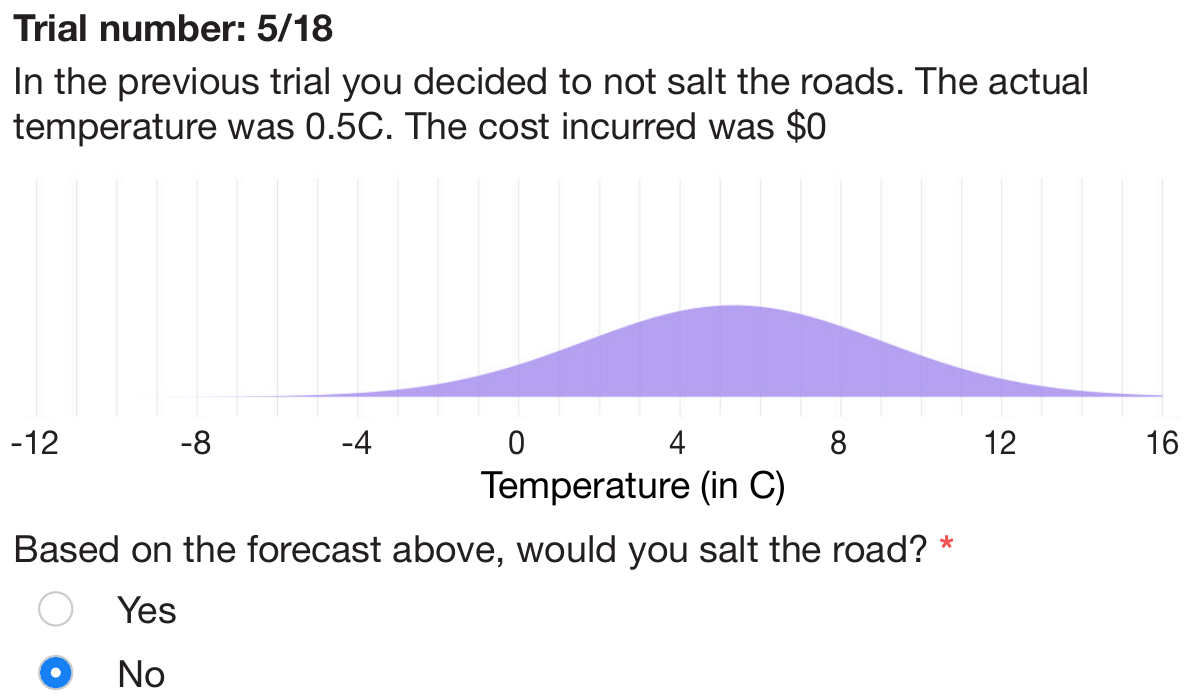}
    \vspace{-6pt}
    \caption{Example of a stimulus seen by a participant in the incentivized density plot condition.}
    \label{fig:density-example}
    \vspace{-12pt}
\end{figure}

\autoref{fig:density-example} shows a screenshot of the experimental interface. The study is available to browse \ourhref{https://abhsarma.github.io/adversarial-effects-experiment/}{here} and the source code for the study is available on \ourhref{https://github.com/abhsarma/adversarial-effects-experiment}{GitHub}.

\customsubsubsection{Experimental Task:} For this study, we adapted the scenario that was used by Joslyn and LeClerc~\cite{joslyn_uncertainty_2012}, where participants were presented with the following hypothetical scenario:

\vspace{-6pt}
\begin{quote}
Assume that you are an analyst of a road maintenance company contracted to treat the roads with salt brine to prevent icing in a U.S. town impacted by severe cold. Applying salt brine to the roads is costly for your company and is also detrimental to the environment, as it can pollute groundwater and kill roadside vegetation. However, not salting the roads can cause significant accidents during freezing temperatures, the costs of which are borne by your company. Your job is to salt the roads when temperatures drop below 0\textdegree C (32\textdegree F). You have a budget of \$18,000 for 18 days. Salting all the roads in the town costs \$1,000 (per night). If you fail to salt the roads and the temperature drops below 0\textdegree C (32\textdegree F), it will cost \$5,000 from your budget. You will be shown a night-time temperature forecast distribution based on which, you will have to decide whether to salt the roads.
\end{quote}
\vspace{-6pt}

\noindent The payoff matrix for the decision problem described above can be represented using the following table:

{
\centering
\vspace{2pt}
\begin{tabular}{Sc|Sc|Sc}
    & $s_1: \text{T} \leq 0$\textdegree C & $s_2: \text{T} > 0$\textdegree C  \\
    \hline
    $a_1: \text{salt}$ & -1000 & -1000 \\
    \hline
    $a_2: \neg \text{salt}$ & -5000 & 0
\end{tabular}
\par%
\vspace{2pt}
}

\noindent Based on this incentive scheme, a decision-maker who wants to maximize expected utility should prefer the action $a_1$ (to salt) if and only if the probability of temperature being below freezing, $\mathrm{Pr(T \leq 0)} = p \geq 0.2$. Thus, $p = 0.2$ is the optimal \textit{crossover point}: the probability at which a decision maker should not have a preference between actions.

\customsubsubsection{Experimental Conditions:} There were three experimental variables in this study: (1) the visualization (interval or density plot); (2) the incentive scheme (advertised as either a participation fee of \$3 or a participation fee of \$1.5 and a bonus of \$0.5 for each \$1,000 remaining in their account at the end of the experiment); (3) the actual forecasted probability of freezing (9 levels, each repeated twice). The experiment used a mixed-factorial design with visualization and incentive scheme varied between-subjects and the forecast varied within subjects.

\customsubsubsection{Procedure:} The experiment consisted of 18 trials and two attention check questions which are interspersed among the 18 trials. In each trial participants are presented with a single forecast in the form of a Normal distribution (see \autoref{fig:density-example}). We varied the means and standard deviations of the Normal distribution to generate different forecasts with different probabilities of freezing: $\mathrm{Pr}(T\leq0\text{\textdegree C}) =$ \{0.595, 0.5, 0.405, 0.315, 0.235, 0.168, 0.115, 0.075, 0.046\}. The attention check questions showed extreme forecasts ($Normal(-11, 0.4)$ and $Normal(11, 0.4)$). The order of the trials were completely randomized. After completing all the trials, participants were asked open-ended questions similar to experiment 1.

\customsubsubsection{Tutorials and Training:} We provided instructions to participants on how to correctly interpret the interval and density plots. This is followed by participants completing four training trials where participants were shown four different temperature forecasts and were asked to report the probability of the temperature falling below a certain marked temperature. Participants were then provided feedback on whether they answered the questions correctly.

\customsubsubsection{Participants:} We recruited all participants from Prolific. Our experiment was only eligible to participants who were fluent in English, and on desktop devices. As per our pre-registrations, we aimed to recruit 280 participants (70 participants in each condition). We experienced some data-logging issues which meant the data for 31 participants were not recorded. After excluding participants who failed to meet our pre-registered attention check criteria (six), we had 243 participants (60 in the \baseci{}, 62 in \incci, 59 in \basedens, and 62 in \incdens). We received explicit consent from all participants to collect and share their responses. The median completion time for participants in the baseline condition was approximately 16 mins, and they were compensated \$3.75 (approximately \$14/h); the median completion time for participants in the incentivized condition was approximately 24 mins, and they received a guaranteed amount of \$4.2 (adjusted up from \$1.5) and the average bonus was \$1.6 (corresponding to an average wage of \$14.5/h).

\subsection{Model Specification}

We used a linear-in-log-odds (llo) model~\cite{zhang_ubiquitous_2012}, which is a good fit to account for the ``distortion of judgement or misperception of probabilities'' \cite{yang_subjective_2023} in such decision-making under uncertainty tasks. This model, which has been used in prior work to model the same task~\cite{sarma_more_2025}, is derived directly from the utility-optimal decision criterion by first translating it into log-odds and then applying a linear transformation to derive a model for participants' decisions:\\

\vspace{-4pt}
\begin{minipage}[h]{\columnwidth}
{\hspace{-12pt}\includegraphics{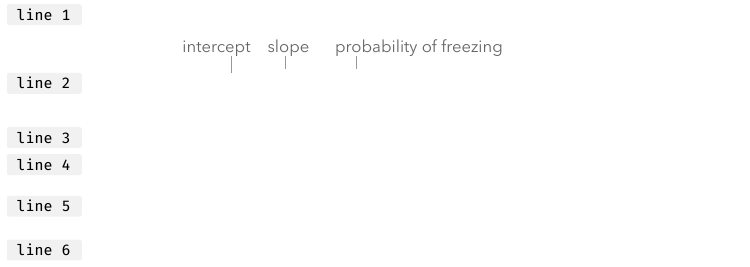}\vspace{-137pt}}
\begin{align*}
    & \mathit{salt} \hskip0.5em \sim \hskip0.5em \text{Binomial}(p_{\textsc{salt}}) \\ \\[6pt]
    & \mathrm{logit}(p_\textsc{salt}) = \alpha_{i} \hskip0.5em + \hskip0.5em \beta_{i} \cdot [ \mathrm{logit}(p) - \mathrm{logit}(0.2)] \\ \\
    & \alpha_{i} \hskip0.5em = \hskip0.5em \alpha_0 + \alpha_{\textsc{vis}[i]} + \alpha_{\textsc{inc}[i]} + \delta_{\alpha, i} \\
    & \beta_{i}  \hskip0.5em = \hskip0.5em \beta_0 \hskip0.12em + \beta_{\textsc{vis}[i]} \hskip0.12em + \beta_{\textsc{inc}[i]} \hskip0.12em + \delta_{\beta, i} \\
    &\onedmatrix{\delta_{\alpha,i}}{\delta_{\beta, i}} \hskip0.5em \sim \hskip0.5em \text{MVNormal}\left(\onedmatrix{0}{0}, \Sigma\right) \\
    & i \in \{1...N\} \hskip4em (N \; participants) \\
\end{align*}
\end{minipage}
\vspace{-12pt}

\customsubsubsection{Line 1:} The binary decision (where 1 represents the decision to salt and 0 represents the decision to not salt) made by a participant in each trial is modeled as a Binomial distribution with probability $p_\textsc{salt}$. 

\customsubsubsection{Line 2:} We assumed that participants' decisions would be a function of the temperature forecast distribution shown, which can be represented using the corresponding probability of freezing value ($p$). The intercept ($\alpha$) parameter controls the fixed point of the function---how people map the optimal crossover point ($p = 0.2$) to the probability of salting---which shifts the crossover point. The further $\alpha$ is from 0, the more bias there is, with negative values of $\alpha$ suggesting that crossover point for the average participant is greater than the optimal, and positive values of $\alpha$ suggesting that crossover point for the average participant is less than the optimal. The slope ($\beta$) parameter controls the degree of sensitivity. The further it is from 1, the more sensitive a participants is to their subjective crossover points.

\customsubsubsection{Lines 3-5:} We expect the intercept ($\alpha_{i}$) and the slope ($\beta_{i}$) parameters to vary between participants, as different participants will likely have different decision-making capacities. $\alpha$ and $\beta$ are the slope parameters for the \textit{average} participant ($\delta_{\alpha, i} = 0$; $\delta_{\beta, i} = 0$), whereas $\delta_{\alpha, i}$ and $\delta_{\beta, i}$ capture differences between each participants' intercepts and slopes compared to the average participant, as random effects.

\customsubsubsection{Priors:} Perfectly unbiased responses would yield values of $\alpha_{i} = 0$. $\beta_{i} = 1$. We thus use priors centered on these values but which also permit the possibility of significant distortion: $\alpha \sim \mathrm{Normal}(0, 1)$ and $\beta \sim \mathrm{Normal}(1, 1)$. We use zero-centered priors for the random effects parameters ($\delta_{\alpha, i}$ and $\delta_{\beta, i}$).

\customsubsubsection{Implementation:} These steps were almost identical to study 1.

\begin{figure}[b!]
    \vspace{-14pt}
    \centering
    \includegraphics[width=0.98\columnwidth]{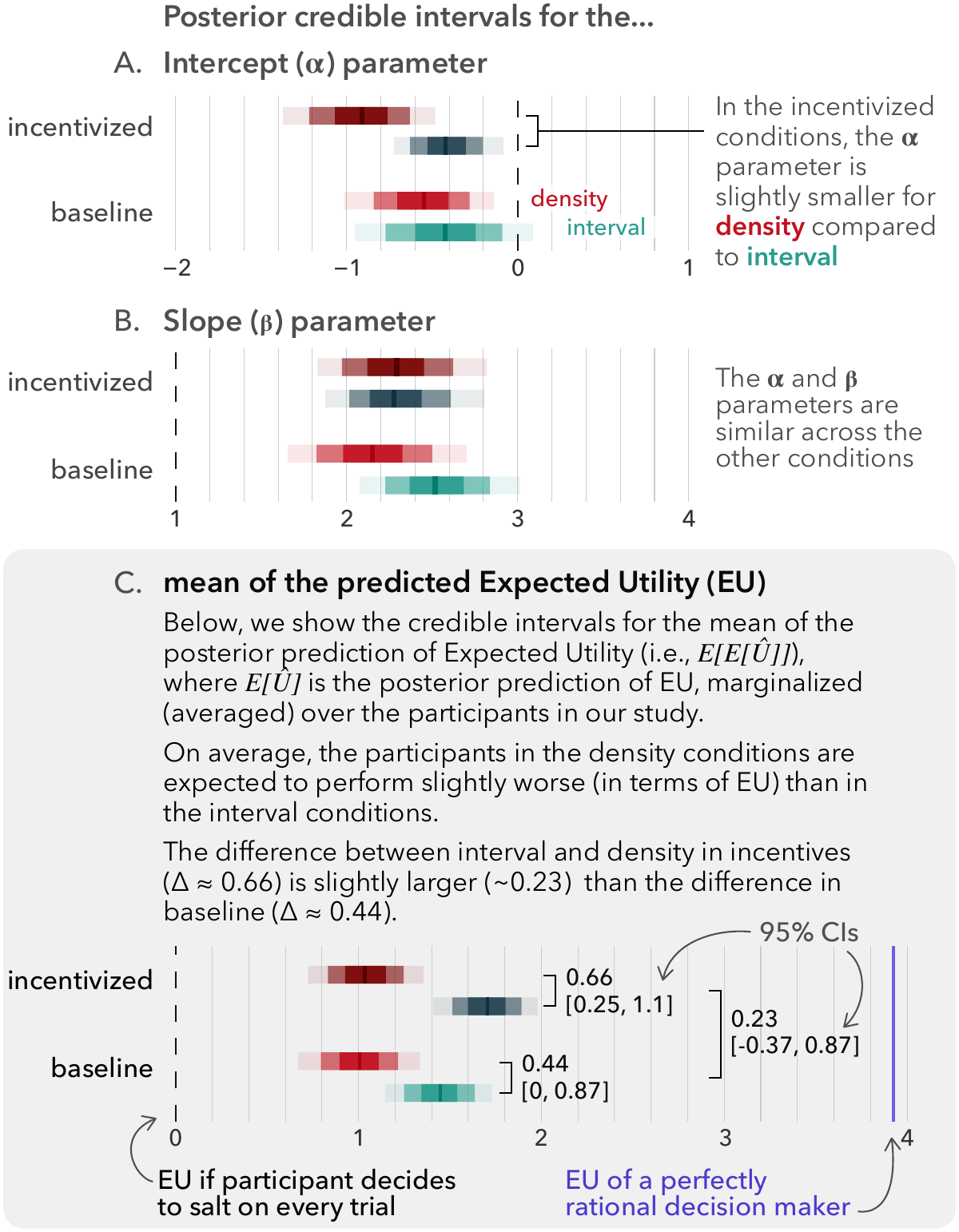}
    \vspace{-6pt}
    \caption{The main result of Experiment 2. We show the posterior credible intervals of $\alpha$, $\beta$ and the mean expected utility for both visual representations across the incentivized and baseline conditions.}
    \label{fig:dm-results}
    \vspace{-4pt}
\end{figure}

\subsection{Results}

The results for Experiment 2 are shown in \autoref{fig:dm-results}. We found the value of $\alpha$ to be negative across all conditions (\autoref{fig:dm-results}A). %
This suggests that the crossover point for the average participant is greater than 0.2, exhibiting a consistent bias towards risk-seeking behavior. The estimated value of $\alpha$ was smaller for \incdens{} compared to the other three conditions. We also found the value of $\beta$ to be significantly larger than one across all conditions (\autoref{fig:dm-results}B)%
, indicating that participants are quite sensitive to the stimuli and their subjective crossover point. The similarity in the estimates of the $\alpha$ and $\beta$ parameters across all conditions suggest that the average participant in every condition likely exhibited very similar decision-making behavior. 

We can directly compare decision quality by estimating the expected utility that the average participant in each of these conditions would achieve, which is a measure of how participants performed based on the information that was provided to them regarding how their performance will be evaluated; we report this measure in \autoref{fig:dm-results}C. Contrary to our apriori expectations, we \textbf{find no evidence to suggest that the average participant performs better in the incentivized conditions} in terms of their predicted expected utility (\incci{: 2.76, [2.31, 3.12]} and \incdens{: 2.33, [1.65, 2.86]}) compared to the baseline conditions (\baseci{: 2.69 [2.19, 3.04]} and \basedens{: 2.35 [1.70, 2.86]}).\footnote{All of these expected utility estimates are in thousands of dollars} 

\begin{figure}[b!]
    \vspace{-16pt}
    \includegraphics[width=\columnwidth]{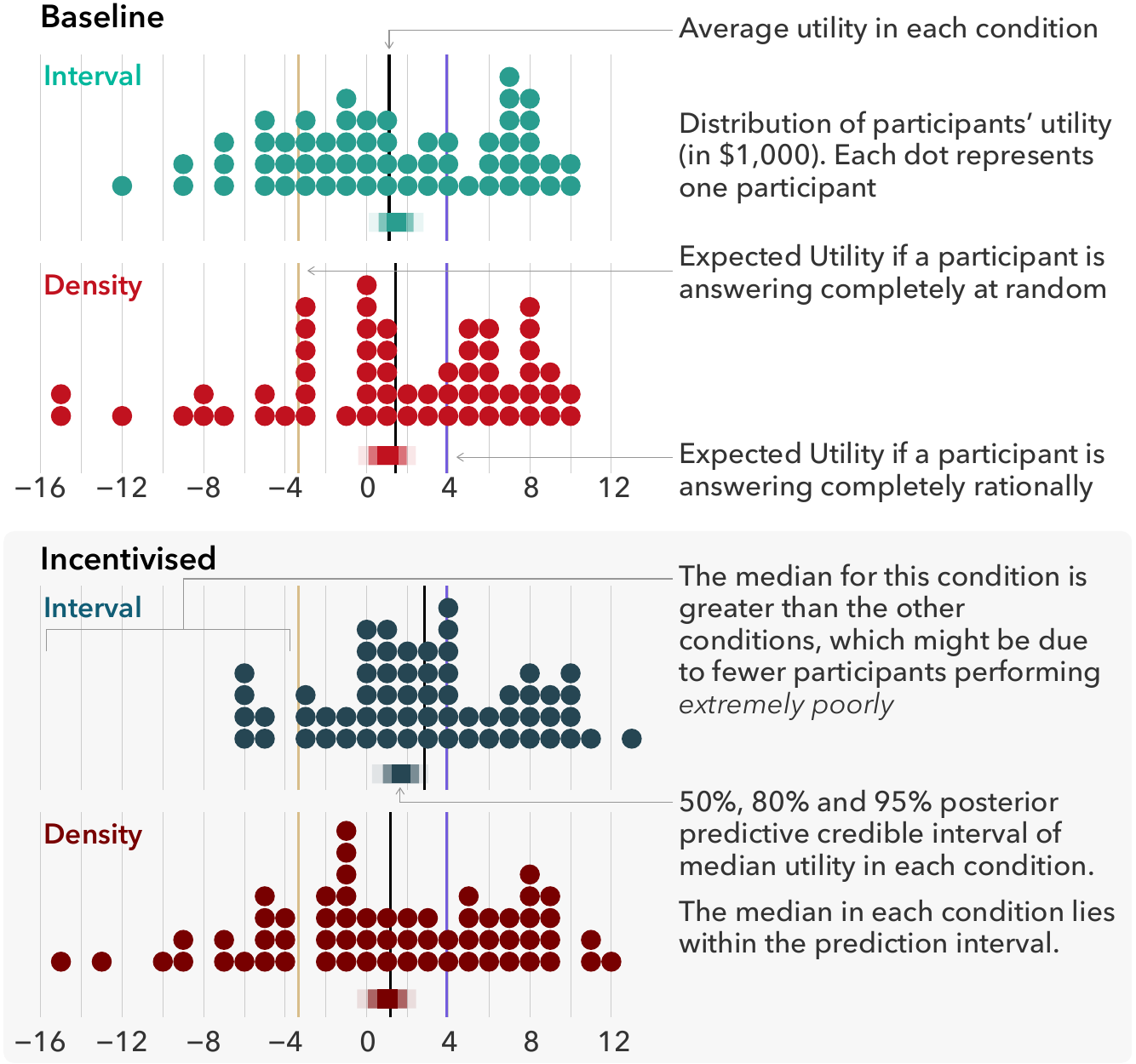}
    \vspace{-16pt}
    \caption{The distribution of utility obtained by the participants and the \textit{predicted} expected utilities on the exact data shown to participants, but with the simulated temperature calculated using the \textsc{rng} in \texttt{R}. The observed median utility for all conditions lie within the prediction intervals.}
    \label{fig:dist-utility}
\end{figure}

We also \textbf{found that the average participant in the interval condition performed slightly better than the average participant in the density condition}, regardless of whether they are incentivized or not. This difference is relatively small---the difference in expected utility is 0.66 95\% CI: [0.25, 1.1] when incentivised and 0.44 95\%CI: [0, 0.87] when not incentivized (see \autoref{fig:dm-results}C). \autoref{fig:dist-utility} shows the distribution of utility attained by participants in the various conditions, as well as the posterior predictive intervals 
(see also \appendixref{\autoref{appendix:exp2}}{Appendix A} for more details). 
The \textit{actual} average utility of participants in the interval condition under incentives was much larger than the other conditions; we suspect that this might have been an artifact of the \textsc{rng} in JavaScript---we implemented a series of checks but could not determine any systemic issues with the \textsc{rng} (\supplement{} $\blacktriangleright$ \texttt{R} $\blacktriangleright$ \texttt{dm-05-rng\_check.qmd}). 
Our finding---that intervals perform better than densities---is in contrast to prior work~\cite{fernandes_uncertainty_2018, kale_visual_2021}, which have typically found that participants make better decisions when information is presented to them using density plots compared to interval plots.%
We discuss the implications of this finding in \autoref{sec:discussion-inconsistent-results}.

As in the first study, we found that participants took slightly longer to complete the task in the incentivized conditions compared to the baseline conditions. \autoref{fig:time-taken}B, visualizes the bootstrapped median and 95\% quantile intervals for the median estimate. The difference in time spent is approximately 80s for intervals (\baseci{: 187s, [159s, 226s]}, \incci{: 266s, [221s, 355s]}), and approximately 32s for density  (\basedens{: 173s, [143s, 187s]}, \incdens{: 205s, [171s, 233s]}). This increase of 20-40\% is quite large, and might suggest that in tasks such as decision-making under uncertainty the use of incentives may lead to more time spent on the task, and \textit{perhaps} more effort induced (depending on how one views the relationship between time spent on an online experiment and effort).

\section{Crowdworkers' Perspectives on Incentives}
\label{sec:qual-responses}
For both studies, we elicited qualitative responses from participants regarding (i) their prior experiences with performance-based incentives, and (ii) how they might have behaved if they were in the other condition (i.e., asked people in the incentivized condition how they would behave if they were given a flat amount and vice-versa). We conducted an exploratory qualitative analysis of participants’ responses using an inductive coding approach to identify a set of codes, which was then used to code all the responses. Below, we report the numbers combined across both studies (N = 444).

Of the responses which we were able to code (224), significantly more participants (115) in our sample claimed to have positive attitudes towards incentives based on their past experience (e.g., ``[...] they can make the task more engaging and encourage more careful decision-making'' or ``Surprised and happy. Its like a little gift you receive because you're doing a good job''), compared to those who explicitly reported mixed or negative feelings (28) towards incentives (e.g., ``Very mixed because most times there is little actual chance of receiving any bonus'' or ``They are often misleading, or sometimes held as a raffle, which is a scammy practice''). Among those who reported positive feelings, some say that it encourages them to do well (34) or makes the task more interesting and engaging (19). On the other hand, those who claim to have negative experience usually believe that it is because they are unfair or conflate bonuses with lotteries or raffles.

For our analysis on how participants self-reported they would behave in the other incentivized condition, we focused only those who were in the incentivized condition as it would be easier for them to imagine the counterfactual. Of the responses we were able to code (192/224), a majority (119) claimed that would have behaved in the same way, there were several participants (35) who believed that the in the absence of incentives they might perform worse because they might be less motivated or focused (e.g., ``If I had been guaranteed [...], my performance probably would have been slightly worse because the bonus made me pay closer attention to the tradeoff between salting costs and freezing risk.''). In the decision-making study, several (12) reported that they would be more risk seeking (e.g., ``[...] would have been slightly less cautious and less incentive-driven.''). For these participants, incentives appear to be having the intended effect. 

However, there were also a number of participants (13) who believed that they would have performed better in the non-incentivized condition perhaps because it would be less stressful (e.g., ``[...] guaranteed payment offers greater financial security and could reduce performance anxiety.'' and ``I think it would be better, as [...] one would treat the study properly because he would get a proper compensation in return''). This might suggest that, because of how incentivized experiments are designed with a lower base pay, for a number of participants, incentives might be having the opposite of the intended effect. In the decision-making study, several participants (16) stated that they might have been less risk averse as they would not be concerned with saving money (e.g., ``With a guaranteed reward, the incentive to save money would vanish. My performance would likely have become highly risk-averse''), again suggesting that incentives may not be having the intended effect for many participants.


\section{Discussion}

\subsection{What is the Effect of Incentives?}
Our work was motivated by several recent studies which have employed incentivized experiments to study decision-making under uncertainty~\cite[e.g.,][]{fernandes_uncertainty_2018, kale_visual_2021, padilla_uncertain_2021, sarma_odds_2024}. The rationale was that decision-making in the real-world entails consequences and incentives would bring that notion of realism to virtual survey world. Based on the \textit{capital-labor-production} theory, our a priori hypotheses were that: (i) the effect of incentives would likely be small, if any, in the perception of correlation as the effort required (or production requirement) is low; and (ii) performance-based incentives would likely lead to a meaningful improvement in performance (in terms of effect size) in the decision-making task as more effort is required to properly perform the task (i.e., higher production requirements). The results of our two experiments (see \Autoref{fig:jnd-results, fig:dm-results}) suggests that, compared to a baseline of providing people a flat rate for participation, introducing performance-based financial incentives may not lead to improved performance in either the perception of correlation task or the decision-making task. 

While expected for the perception of correlation task, this unexpected violation of our apriori expectations could potentially be explained by the capital-labor-production theory---it is possible that the the decision-making task may have been too difficult. We observed that participants in the incentivized conditions did spend longer in completing the task (see \autoref{fig:time-taken}B), and this difference in time spent was particularly large for the decision-making task. \textit{If} time spent on a task is considered a reasonable proxy for effort exerted by a participant, this result suggests that the average participant in the incentivized condition is likely putting in more effort then the average participant in the non-incentivized condition, even though they are not necessarily performing better on the task. We discuss potential implications for studying decision quality in \autoref{sec:discussion-ecological-validity}.

Prior work~\cite{camerer_effects_1999} also suggests that incentives often reduce variance in responses, even if they do not have an impact on mean performance. Our model allows us to estimate this variance, and while the variance parameters estimated by our model, for both studies, are smaller for the incentivized condition compared to non-incentivized conditions, these differences appear to be very small (see the ``Effect on Variance'' subsection in \supplement{} $\blacktriangleright$ R $\blacktriangleright$ \texttt{dm-04-analysis.qmd}).

\subsection{Why Were Our Results Inconsistent With Prior Uncertainty Visualization Studies?}
\label{sec:discussion-inconsistent-results}

Perhaps the most surprising aspect of our experimental results was that the average participant making decisions using interval plots was expected to perform marginally better than with density plots. This is contrast to prior work~\cite[e.g.,][]{kale_visual_2021, fernandes_uncertainty_2018, sarma_odds_2024} which has typically found that, in incentivized experiments, participants make better decisions when the uncertainty distribution is represented using densities rather than 95\% interval plots, as densities provide complete distributional information. We expected the same result to hold for our study, even though we varied our design of interval plots. 

\begin{figure}[t!]
    \centering
    \includegraphics[width=0.95\columnwidth]{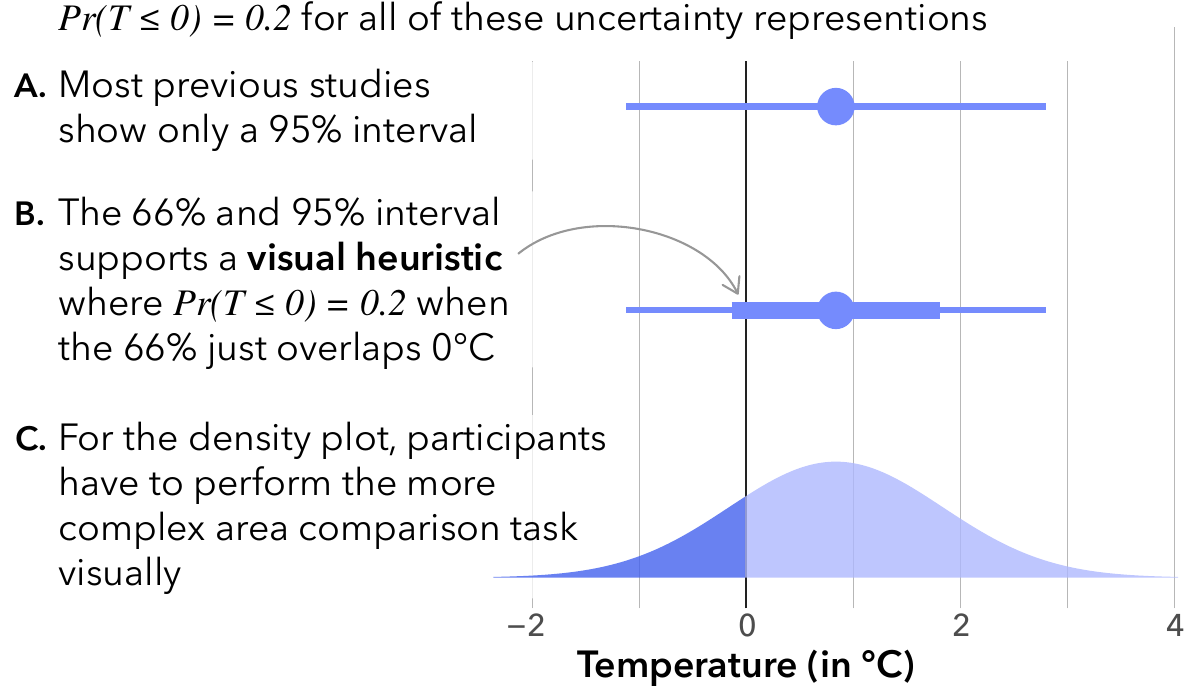}
    \vspace{-10pt}
    \caption{Different uncertainty visualizations used in prior work (A, C)~\cite[e.g.,][]{kale_visual_2021, fernandes_uncertainty_2018, sarma_odds_2024}, and in our work (B, C). 
    }
    \label{fig:visual-heuristic}
    \vspace{-18pt}
\end{figure}

We initially suspected that it was because, unlike previous experiments, we presented participants with both the 66\% and 95\% intervals. The two intervals provide additional information relative to prior studies, and supports a visual heuristic to make utility optimal decisions---if the 66\% interval is slightly overlapping with the 0\textdegree C line, this indicates an approximately 20\% probability that the temperature is going to be below freezing~\autoref{fig:visual-heuristic}. In a ``just 95\% interval'' condition, as in prior work, there is no such easy visual heuristic that a participant can rely on. In contrast, a participant in the density condition would have to compare areas to make the utility optimal decision, which might arguably have been more difficult, given how poor people are at area perception~\cite{stevens_psychophysical_1957}. However, a follow-up study (see \appendixref{\autoref{appendix:exp3}}{Appendix B}) comparing 95\% intervals with density plots again showed no difference in participants' decision quality between the two conditions. One possible cause could be the probability distributions used as stimuli in our study---the distributions were of varying standard deviations, which can interfere with the visual heuristics that participants use to estimate probabilities from densities~\cite{fygenson_impact_2025, fygenson_croissant_2026}. Our results paint a conflicting picture regarding the effectiveness of density plots for uncertainty communication, and warrants further research.

\subsection{Ecological Validity of Decision-Making Studies}
\label{sec:discussion-ecological-validity}

Given the results of our study, and how they were contrary to our initial hypotheses, we did consider shelving this paper. However, we realized that this would be a terrible idea for two reasons: (i) we would be contributing to the file drawer problem~\cite{rosenthal_file_1979, franco_publication_2014} (self-selecting to not publish studies with negative results), and (ii) even though our study does not find evidence for the effect of incentives, it does raise some interesting questions about the use of incentivized experiments for studying decision quality and the ecological validity of studies run on crowd-sourcing platforms such as Prolific.


\customsubsubsection{Why Do We Conduct Incentivized Decision-Making Studies?}\\
Early work in uncertainty visualization compared different representations based on subjective measures such as preference or intuitiveness (e.g., which representation do you prefer?)~\cite{maceachren_visual_2012, dong_uncertainty_2017, greis_uncertainty_2018}, ease of use~\cite{retchless_guidance_2016}, and subjective probability (e.g., on a scale of 1-7, how likely do you think the event is going to occur?) or confidence (e.g., on a scale of 1-7, how confident are you about your prediction?)~\cite{correll_error_2014, ferreira_sample-oriented_2014, greis_uncertainty_2018, kay_when_2016}. A critique of these types of studies is the lack of an objective benchmark: a user might perceive visualization A as being more intuitive than visualization B, or they might experience a greater or lesser degree of subjective uncertainty when viewing it. But how do we know that they are perceiving the \textit{correct} level of uncertainty? Is perceiving less uncertainty or more uncertainty better?

Other studies have investigated participants' ability to accurately estimate the probability of an event occuring using uncertainty visualizations (which sometimes reduces to visually estimating tail probabilities)~\cite[e.g.,][]{hullman_hypothetical_2015, kay_when_2016, correll_value-suppressing_2018}. These studies include an objective benchmark---accuracy of probability estimation---which could be used to evaluate participants' performance. However, even if people are able to read and report numerical probabilities accurately, it is possible that their actual behavior diverges from what would be expected based on the reported probabilities, either due to bias or error, or because they associate different, subjective probabilities to uncertainty~\cite{loewenstein_risk_2001}. As such, measuring uncertainty through accuracy in reading numerical probability values may not necessarily tell us how good people are at using those probabilities to make good decisions.

Since uncertainty visualizations are typically used in the real world to make decisions and a person in the real-world is incentivized to make a correct decision because it has real consequences (e.g., we consult the weather forecast not because we care about knowing the probability of rain, but because we need to decide whether we should carry an umbrella or not), studying decision quality using uncertainty visualizations represented, arguably, a more ecological valid approach of evaluating different uncertainty communication approaches. While other metrics for measuring the effect of uncertainty visualizations could tell us that two visualizations are \textit{different}, they do not necessarily tell us which one is \textit{better}. Asking people to perform incentivized decision tasks, \textit{in theory}, allowed us to measure decision quality against an objective baseline---\textit{the expected utility of a rational decision-maker}---by comparing how far are people from that baseline.

Yet there are a couple of assumptions implicit in this line of reasoning: that all participants in incentivized decision-making tasks are attempting to maximize expected utility; and (assuming they are trying to maximize expected utility) people are able to translate the provided incentive structure or cost function into a decision rule based on probabilities (which is how the uncertainty information is typically presented to them). If these two assumptions hold, incentivized decision-making style experiments can be considered more ecologically valid than other approaches for evaluating uncertainty visualizations. Under these conditions, a decision-making study of uncertainty visualizations measures how people make decisions based on their subjective representation of uncertainty or subjective risk tolerance threshold, and an uncertainty visualization can be considered more effective if they allow a user to make decisions which are close to utility optimal. However, the qualitative analysis of participants responses to the open-ended questions in our study, and some introspective reflection, make us question these assumptions.


\customsubsubsection{Are Participants Trying to Maximize Expected Utility?}\\
In their qualitative responses, a few participants stated to have adopted a risk-seeking strategy, as indicated by statements such as ``gambling with [a] small amount is fun and there is no real concern of losing money'' or ``there is less personal `skin in the game,' so I might have been more willing to gamble on borderline forecasts.'' We also observed something similar in prior work~\cite{sarma_odds_2024}, where some participants stated that they started taking risks after their initial decision-making strategy did not appear to have been paying off. While a few errant participants are to be expected, and can be easily accounted for in a mixed-effects regression model like the one we used, it is unclear how prevalent these attitudes are. In addition, our qualitative analysis suggests that (\autoref{sec:qual-responses}), due to the competing goals of crowdworkers on Prolific, incentives can have the opposite of the desired effect---as the guaranteed payment was low, instead of trying to maximize expected utility, participants simply behaved as if they were being paid the small base pay. If a significant proportion of participants are not trying to maximize expected utility and instead adopting a risk-seeking attitude when performing decision-making tasks, we might \textbf{need to evaluate how incentive structures in experiments are designed so that they elicit the behavior we \textit{normatively} want}.

\customsubsubsection{Can Participants Maximize Expected Utility?}\\
Even if participants were trying to maximize expected utility, it is possible that they were not able to translate this to a decision rule with the optimal crossover point ($p = 0.2$). Of the participants for whom we could deduce a crossover point (53), 25 participants stated that they decided to salt the roads when the chances of freezing were around 30\% or higher (the crossover point for the remainder were at or around 20\%), suggesting an imperfect translation from the given cost function to a decision rule.\footnote{We also observed a very colorful (mis)interpretation of probability, not too dissimilar to the misinterpretations of what \textit{probability of rain} means as described by Gigerenzer et al.~\cite{gigerenzer_30_2005}: ``If more than 50\% of the night, or a significant proportion, was below zero, I salted the roads. If there were more hours above 0, I didn't salt because it wasn't necessary''} 
This is reflected in the model estimates of $\alpha$ (\autoref{fig:dm-results}A), and is consistent with findings from prior work on a similar task~\cite{sarma_more_2025}.
There is a qualitative difference between risk-seeking behavior when a participant decides to salt when $p = 0.3$ knowing that the optimal crossover point is $p = 0.2$, and ``risk-seeking behavior'' when a participant decides to salt when $p = 0.3$ believing that the optimal crossover point is $p = 0.3$. When evaluating decision quality, we want to make sure we are measuring the former and not the latter. 

A different (mis)calculation can be observed among participants who stated that they would be more risk averse in the baseline conditions as they would not be incentivized to ``save money.'' If some participants did not know how to maximize expected utility, it might also mean that they are not sensitive to the incentive structure that is provided to them (i.e., could participants behave similarly if the penalty for not salting was \$3,000 instead of \$5,000?). From the current study, it is unclear whether some participants' inability to identify a decision rule is simply an artifact of this specific task, the format of presenting incentive structures, difficulties in interpreting single event probabilities properly~\cite{gigerenzer_30_2005}, not being sufficiently sensitive to incentives or something else entirely, and therefore warrants further research. 


\customsubsubsection{Other Possibly Important Factors To Consider.}\\
In Experiment 2, we provided feedback to participants after every trial regarding what the simulated temperature was, and the costs incurred. It is possible that this immediate feedback could have led participants to adopt a sub-optimal strategy---according to Achtziger et al.~\cite{achtziger_higher_2015}, ``the human tendency to repeat successful actions and avoid those which led to failure can impair performance by focusing attention on win/lose outcomes and away from the probabilities of the relevant uncertain events.'' While the setup of  the decision-making task used in our study does not allow feedback to be withheld, it is worth exploring whether participants are more rational decision-makers under uncertainty for tasks where immediate feedback is not provided.

The use of incentives assumes that participants who are given a flat payment regardless of how they perform will put in less effort, and overlooks the fact that many participants in the non-incentivised conditions might have strong intrinsic motivation to provide high-quality data and perform well in the study~\cite{cerasoli_intrinsic_2014}. According to Camerer and Hogarth~\cite{camerer_effects_1999}, ``some people like mental effort, and those people disproportionately volunteer for experiments!'' If this attitude is widespread among crowdworkers on platforms such as Prolific, then the additional effort for implementing incentives might not be worthwhile for most researchers. An additional aspect to consider here are the policies of Prolific as a platform---the quality checks implemented by Prolific could serve as another source of motivation for crowdworkers to provide high quality responses in experiments, reducing the potential effect of performance-based incentives.

Finally, participants in our non-incentivized conditions were still informed on whether they answered a question correctly and how many they answered correctly so far (perception of correlation), or how much budget they have remaining (decision-making). In other words, while participants were not financially rewarded for their performance, they could interpret these as virtual rewards. This was a conscious choice in our experiment in order to minimize the discrepancies between the two experimental conditions. However, it is possible that, in crowdsourcing platforms such as Prolific, if participants are already highly intrinsically motivated to do well, virtual rewards could have a similar effect on effort as real rewards.


 


\subsection{The Possible Role of Incentives in Other Studies}
Our work was an initial investigation into the role of incentives in two prototypical types of visualization studies. While our results may not generalize to all visualization tasks, we can speculate on how incentives might impact certain other types of visualization tasks, based on our findings and the \textit{capital-labor-production} theory~\cite{camerer_effects_1999}. Studies on graphical perception~\cite[e.g.,][]{cleveland_graphical_1984, davis_risks_2024, heer_crowdsourcing_2010} or color perception \cite[e.g.][]{szafir_modeling_2018, reda_graphical_2018, reda_rainbow_2023, reda_rainbows_2021}, which require users to perform similarly low-level perceptual tasks as Study 1 are likely not going to benefit from the increased effort than incentives induce. 

There are likely a class of visualization tasks such as some matrix-based tasks (e.g., Nobre et al.~\cite{nobre_evaluating_2020}, performed a crowdsourced evaluation of node-link diagrams and adjacency matrices where participants were required to undergo lots of training and perform complex tasks), which either require a lot of a priori knowledge (cognitive capital) or a lot of effort (high production requirements), and thus may not benefit from incentives (and decision-making under uncertainty could possibly fall into this category as well!). In other words, these represent tasks where participants may perform poorly not because they do not want to put in more effort, but rather because they are not sure of how to perform the task well and putting in more effort may not necessarily help them perform the task better.

The type of tasks that we believe may benefit from incentives are ones which do not require a lot of cognitive capital but can feel a bit tedious; incentives might motivate participants to put in the required additional effort to complete the task accurately. An example could be the \textit{finding the shortest path between two nodes} task using node-link or adjacency matrix representations---node-link diagrams have been found to be more accurate~\cite{okoe_node-link_2019, keller_matrices_2006, ghoniem_comparison_2004}, but if incentivized, participants may be more willing to put in the effort to perform the tedious steps involved for adjacency matrices. A challenge here is ensuring that the incentive function is well designed and not arbitrary~\cite{wu_rational_2024, kale_logic_2025}. We hope to explore such tasks in future work.

\subsection{Ethical Considerations for Incentivized Experiments}
In both experiments, participants in the incentivized conditions took longer to complete the task compared to participants in the baseline condition. Therefore, (i) the initial median hourly wage for participants was lower than the advertised amount, and, after factoring in bonus, payment was below our target fair wage of \$15/h; and (ii) the use of performance-based financial incentives increased the variance in wages that participants received---a participant who can perform the task quickly and well can receive a significantly higher wage than a participant who performs it slowly and poorly. 

Ensuring a fair median wage for participants is  still possible for incentivized experiments, as the researcher can increase the guaranteed amount so that the median wage, taking into account bonuses, reflects what is considered a fair wage. In order to ensure a wage of \$15/h, and to comply with Prolific's minimum guaranteed wage of \$8/h, we increased the fixed participation fee (by \$0.35 in Experiment 1, and by \$2.7 in Experiment 2). Addressing the issue of variance in wages, however, is significantly more complicated, and to some degree not compatible with the goal of performance-based incentives. While there is always going to be some degree of variance in compensation in crowdsourced studies, as not every participant takes the same amount of time, using bonuses as a significant portion of the compensation exacerbates this issue. A solution here might be to increase the guaranteed amount to ensure that participants receive a fair hourly wage. However, as research teams typically operate under considerable budget constraints, this would likely mean that they would have to in turn reduce the bonus amounts that participants can earn, making them less ``incentivizing.'' 

Running incentivized studies has real costs on the experimenter---increased implementation and design burden while setting up the experiment, the need to manually compensate participants on the crowdsourcing platform, the seemingly low advertised compensation on the crowdsourcing platform (which can have reputational costs for future studies). Taking all of this into account, there is a credible argument to be made against the use of performance-based financial incentives due to the high costs, seemingly low rewards, and the associated ethical questions regarding fair labour practices.

\section{Conclusion}

We conducted two studies to test the value of performance-based monetary incentives for two different types analysis questions that commonly surface in visualization studies: a lower-level perceptual task, and a higher-level decision-making task. We did not find evidence to suggest that incentives affect performance on either task. To our knowledge, our work was the first formal study of monetary incentives in crowdsourced visualization studies. While we acknowledge that our results may not easily translate to other scenarios (e.g., lab studies, different types of tasks, significantly higher incentive pay, etc.), they seem to suggest that incentives, as currently deployed in many empirical studies, may matter less as a \textit{tacit factor} than we expected. Given recent calls for incorporating decision theory, utility functions and incentives into visualization and \textsc{hci} experiments~\cite{wu_rational_2024, kale_logic_2025}, our results provide an empirical data point that putting them into practice may not be as straightforward as we might have initially presumed.



\section*{Acknowledgments}

We would like to thank to Jack Wilburn and Zach Cutler for technical support on this project, and the anonymous reviewers for their feedback on the manuscript. This work was partially funded by the National Science Foundation awards 2213756 and 2403094.

\bibliographystyle{abbrv-doi-hyperref}

\bibliography{references}

@article{achtziger_higher_2015,
  title = {Higher Incentives Can Impair Performance: Neural Evidence on Reinforcement and Rationality},
  shorttitle = {Higher Incentives Can Impair Performance},
  author = {Achtziger, Anja and {Al{\'o}s-Ferrer}, Carlos and H{\"u}gelsch{\"a}fer, Sabine and Steinhauser, Marco},
  year = 2015,
  month = nov,
  journal = {Social Cognitive and Affective Neuroscience},
  volume = {10},
  number = {11},
  pages = {1477--1483},
  issn = {1749-5024, 1749-5016},
  doi = {10.1093/scan/nsv036},
  urldate = {2026-06-13},
  copyright = {http://creativecommons.org/licenses/by-nc/4.0/},
  langid = {english}
}

@techreport{cala_financial_2022,
  type = {Working {{Paper}}},
  title = {Financial Incentives and Performance: {{A}} Meta-Analysis of Economics Evidence},
  shorttitle = {Financial Incentives and Performance},
  author = {Cala, Petr and Havr{\'a}nek, Tom{\'a}{\v s} and Havr{\'a}nkov{\'a}, Zuzana and Matousek, Jindrich and Nov{\'a}k, Jiri},
  year = 2022,
  number = {27/2022},
  institution = {IES Working Paper},
  urldate = {2026-08-06},
  copyright = {https://www.econstor.eu/dspace/Nutzungsbedingungen},
  langid = {english}
}

@incollection{camerer_effects_1999,
  title = {The {{Effects}} of {{Financial Incentives}} in {{Experiments}}: {{A Review}} and {{Capital-Labor-Production Framework}}},
  shorttitle = {The {{Effects}} of {{Financial Incentives}} in {{Experiments}}},
  booktitle = {Elicitation of {{Preferences}}},
  author = {Camerer, Colin F. and Hogarth, Robin M. and Budescu, David V. and Eckel, Catherine},
  editor = {Fischhoff, Baruch and Manski, Charles F.},
  year = 1999,
  pages = {7--48},
  publisher = {Springer Netherlands},
  address = {Dordrecht},
  doi = {10.1007/978-94-017-1406-8_2},
  urldate = {2026-06-19},
  isbn = {978-90-481-5776-1 978-94-017-1406-8},
  langid = {english}
}

@article{cerasoli_intrinsic_2014,
  title = {Intrinsic Motivation and Extrinsic Incentives Jointly Predict Performance: {{A}} 40-Year Meta-Analysis.},
  shorttitle = {Intrinsic Motivation and Extrinsic Incentives Jointly Predict Performance},
  author = {Cerasoli, Christopher P. and Nicklin, Jessica M. and Ford, Michael T.},
  year = 2014,
  journal = {Psychological Bulletin},
  volume = {140},
  number = {4},
  pages = {980--1008},
  issn = {1939-1455, 0033-2909},
  doi = {10.1037/a0035661},
  urldate = {2026-06-13},
  langid = {english}
}

@article{cleveland_graphical_1984,
  title = {Graphical {{Perception}}: {{Theory}}, {{Experimentation}}, and {{Application}} to the {{Development}} of {{Graphical Methods}}},
  shorttitle = {Graphical {{Perception}}},
  author = {Cleveland, William S. and McGill, Robert},
  year = 1984,
  month = sep,
  journal = {Journal of the American Statistical Association},
  volume = {79},
  number = {387},
  pages = {531--554},
  issn = {0162-1459, 1537-274X},
  doi = {10.1080/01621459.1984.10478080},
  urldate = {2026-06-19},
  langid = {english}
}

@misc{cmdstanr,
  title = {Cmdstanr: {{R Interface}} to '{{CmdStan}}'},
  author = {Gabry, Jonah and {\v C}e{\v s}novar, Rok and Johnson, Andrew and Bronder, Steve},
  year = 2025
}

@article{correll_error_2014,
  title = {Error {{Bars Considered Harmful}}: {{Exploring Alternate Encodings}} for {{Mean}} and {{Error}}},
  shorttitle = {Error {{Bars Considered Harmful}}},
  author = {Correll, Michael and Gleicher, Michael},
  year = 2014,
  month = dec,
  journal = {IEEE Transactions on Visualization and Computer Graphics},
  volume = {20},
  number = {12},
  pages = {2142--2151},
  issn = {1077-2626},
  doi = {10.1109/TVCG.2014.2346298},
  urldate = {2026-06-19},
  copyright = {https://ieeexplore.ieee.org/Xplorehelp/downloads/license-information/IEEE.html},
  langid = {english}
}

@inproceedings{correll_value-suppressing_2018,
  title = {Value-{{Suppressing Uncertainty Palettes}}},
  booktitle = {Proceedings of the 2018 {{CHI Conference}} on {{Human Factors}} in {{Computing Systems}}},
  author = {Correll, Michael and Moritz, Dominik and Heer, Jeffrey},
  year = 2018,
  month = apr,
  pages = {1--11},
  publisher = {ACM},
  address = {Montreal QC Canada},
  doi = {10.1145/3173574.3174216},
  urldate = {2026-08-06},
  isbn = {978-1-4503-5620-6},
  langid = {english}
}

@article{cutler_revisit_2026,
  title = {{{ReVISit}} 2: {{A Full Experiment Life Cycle User Study Framework}}},
  shorttitle = {{{ReVISit}} 2},
  author = {Cutler, Zach and Wilburn, Jack and Shrestha, Hilson and Ding, Yiren and Bollen, Brian and Nadib, Khandaker Abrar and He, Tingying and McNutt, Andrew and Harrison, Lane and Lex, Alexander},
  year = 2026,
  month = jan,
  journal = {IEEE Transactions on Visualization and Computer Graphics},
  volume = {32},
  number = {1},
  pages = {13--23},
  issn = {1941-0506},
  doi = {10.1109/TVCG.2025.3633896},
  urldate = {2026-07-02}
}

@article{davis_risks_2024,
  title = {The {{Risks}} of {{Ranking}}: {{Revisiting Graphical Perception}} to {{Model Individual Differences}} in {{Visualization Performance}}},
  shorttitle = {The {{Risks}} of {{Ranking}}},
  author = {Davis, Russell and Pu, Xiaoying and Ding, Yiren and Hall, Brian D. and Bonilla, Karen and Feng, Mi and Kay, Matthew and Harrison, Lane},
  year = 2024,
  month = mar,
  journal = {IEEE Transactions on Visualization and Computer Graphics},
  volume = {30},
  number = {3},
  pages = {1756--1771},
  issn = {1077-2626, 1941-0506, 2160-9306},
  doi = {10.1109/TVCG.2022.3226463},
  urldate = {2026-06-19},
  copyright = {https://ieeexplore.ieee.org/Xplorehelp/downloads/license-information/IEEE.html},
  langid = {english}
}

@article{dong_uncertainty_2017,
  title = {Uncertainty {{Visualization}} for {{Mobile}} and {{Wearable Devices Based Activity Recognition Systems}}},
  author = {Dong, Miaomiao and Chen, Ling and Wang, Liwen and Jiang, Xianta and Chen, Gencai},
  year = 2017,
  month = feb,
  journal = {International Journal of Human--Computer Interaction},
  volume = {33},
  number = {2},
  pages = {151--163},
  publisher = {Taylor \& Francis},
  issn = {1044-7318},
  doi = {10.1080/10447318.2016.1224527},
  urldate = {2026-06-19}
}

@inproceedings{dragicevic_increasing_2019,
  title = {Increasing the {{Transparency}} of {{Research Papers}} with {{Explorable Multiverse Analyses}}},
  booktitle = {Proceedings of the 2019 {{CHI Conference}} on {{Human Factors}} in {{Computing Systems}}},
  author = {Dragicevic, Pierre and Jansen, Yvonne and Sarma, Abhraneel and Kay, Matthew and Chevalier, Fanny},
  year = 2019,
  month = may,
  pages = {1--15},
  publisher = {ACM},
  address = {Glasgow Scotland Uk},
  doi = {10.1145/3290605.3300295},
  urldate = {2026-06-19},
  isbn = {978-1-4503-5970-2},
  langid = {english}
}

@inproceedings{fernandes_uncertainty_2018,
  title = {Uncertainty {{Displays Using Quantile Dotplots}} or {{CDFs Improve Transit Decision-Making}}},
  booktitle = {Proceedings of the 2018 {{CHI Conference}} on {{Human Factors}} in {{Computing Systems}}},
  author = {Fernandes, Michael and Walls, Logan and Munson, Sean and Hullman, Jessica and Kay, Matthew},
  year = 2018,
  month = apr,
  pages = {1--12},
  publisher = {ACM},
  address = {Montreal QC Canada},
  doi = {10.1145/3173574.3173718},
  urldate = {2026-06-19},
  isbn = {978-1-4503-5620-6},
  langid = {english}
}

@inproceedings{ferreira_sample-oriented_2014,
  title = {Sample-Oriented Task-Driven Visualizations: Allowing Users to Make Better, More Confident Decisions},
  shorttitle = {Sample-Oriented Task-Driven Visualizations},
  booktitle = {Proceedings of the {{SIGCHI Conference}} on {{Human Factors}} in {{Computing Systems}}},
  author = {Ferreira, Nivan and Fisher, Danyel and Konig, Arnd Christian},
  year = 2014,
  month = apr,
  pages = {571--580},
  publisher = {ACM},
  address = {Toronto Ontario Canada},
  doi = {10.1145/2556288.2557131},
  urldate = {2026-06-19},
  isbn = {978-1-4503-2473-1},
  langid = {english}
}

@article{franco_publication_2014,
  title = {Publication Bias in the Social Sciences: {{Unlocking}} the File Drawer},
  shorttitle = {Publication Bias in the Social Sciences},
  author = {Franco, Annie and Malhotra, Neil and Simonovits, Gabor},
  year = 2014,
  month = sep,
  journal = {Science},
  volume = {345},
  number = {6203},
  pages = {1502--1505},
  publisher = {American Association for the Advancement of Science},
  doi = {10.1126/science.1255484},
  urldate = {2026-06-19}
}

@article{fygenson_croissant_2026,
  title = {Croissant {{Charts}}: {{Modulating}} the {{Performance}} of {{Normal Distribution Visualizations}} with {{Affordances}}},
  shorttitle = {Croissant {{Charts}}},
  author = {Fygenson, R. and Bertini, E. and Padilla, L. M.},
  year = 2026,
  month = nov,
  journal = {Computer Graphics Forum},
  pages = {e70463},
  issn = {1467-8659},
  doi = {10.1111/cgf.70463},
  urldate = {2026-06-15},
  copyright = {\copyright{} 2026 Eurographics - The European Association for Computer Graphics and John Wiley \& Sons Ltd.},
  langid = {english}
}

@article{fygenson_impact_2025,
  title = {Impact of {{Vertical Scaling}} on {{Normal Probability Density Function Plots}}},
  author = {Fygenson, Racquel and Padilla, Lace},
  year = 2025,
  month = jan,
  journal = {IEEE Transactions on Visualization and Computer Graphics},
  volume = {31},
  number = {1},
  pages = {984--994},
  issn = {1941-0506},
  doi = {10.1109/TVCG.2024.3456396},
  urldate = {2026-06-15}
}

@article{galesic_using_2009,
  title = {Using Icon Arrays to Communicate Medical Risks: {{Overcoming}} Low Numeracy.},
  shorttitle = {Using Icon Arrays to Communicate Medical Risks},
  author = {Galesic, Mirta and {Garcia-Retamero}, Rocio and Gigerenzer, Gerd},
  year = 2009,
  journal = {Health Psychology},
  volume = {28},
  number = {2},
  pages = {210--216},
  issn = {1930-7810, 0278-6133},
  doi = {10.1037/a0014474},
  urldate = {2026-06-19},
  langid = {english}
}

@inproceedings{ghoniem_comparison_2004,
  title = {A {{Comparison}} of the {{Readability}} of {{Graphs Using Node-Link}} and {{Matrix-Based Representations}}},
  booktitle = {{{IEEE Symposium}} on {{Information Visualization}}},
  author = {Ghoniem, M. and Fekete, J.-D. and Castagliola, P.},
  year = 2004,
  pages = {17--24},
  publisher = {IEEE},
  address = {Austin, TX, USA},
  doi = {10.1109/INFVIS.2004.1},
  urldate = {2026-06-19},
  langid = {english}
}

@article{gigerenzer_30_2005,
  title = {``{{A}} 30\% {{Chance}} of {{Rain Tomorrow}}'': {{How Does}} the {{Public Understand Probabilistic Weather Forecasts}}?},
  shorttitle = {``{{A}} 30\% {{Chance}} of {{Rain Tomorrow}}''},
  author = {Gigerenzer, Gerd and Hertwig, Ralph and Van Den Broek, Eva and Fasolo, Barbara and Katsikopoulos, Konstantinos V.},
  year = 2005,
  journal = {Risk Analysis},
  volume = {25},
  number = {3},
  pages = {623--629},
  issn = {1539-6924},
  doi = {10.1111/j.1539-6924.2005.00608.x},
  urldate = {2026-06-19},
  langid = {english}
}

@article{gonzalez-rubio_webers_2026,
  title = {Weber's {{Law}} in Walking: Sensory Scaling Is Observed in Multi-Sensory, Dynamic Tasks},
  shorttitle = {Weber's {{Law}} in Walking},
  author = {{Gonzalez-Rubio}, Marcela and Iturralde, Pablo A. and {Torres-Oviedo}, Gelsy},
  year = 2026,
  month = jun,
  journal = {Scientific Reports},
  issn = {2045-2322},
  doi = {10.1038/s41598-026-54948-5},
  urldate = {2026-06-19},
  langid = {english}
}

@inproceedings{greis_uncertainty_2018,
  title = {Uncertainty {{Visualization Influences}} How {{Humans Aggregate Discrepant Information}}},
  booktitle = {Proceedings of the 2018 {{CHI Conference}} on {{Human Factors}} in {{Computing Systems}}},
  author = {Greis, Miriam and Joshi, Aditi and Singer, Ken and Schmidt, Albrecht and Machulla, Tonja},
  year = 2018,
  month = apr,
  pages = {1--12},
  publisher = {ACM},
  address = {Montreal QC Canada},
  doi = {10.1145/3173574.3174079},
  urldate = {2026-06-19},
  isbn = {978-1-4503-5620-6},
  langid = {english}
}

@article{hall_survey_2022,
  title = {A {{Survey}} of {{Tasks}} and {{Visualizations}} in {{Multiverse Analysis Reports}}},
  author = {Hall, Brian D. and Liu, Yang and Jansen, Yvonne and Dragicevic, Pierre and Chevalier, Fanny and Kay, Matthew},
  year = 2022,
  journal = {Computer Graphics Forum},
  volume = {41},
  number = {1},
  pages = {402--426},
  issn = {1467-8659},
  doi = {10.1111/cgf.14443},
  urldate = {2026-06-19},
  langid = {english}
}

@article{harrison_ranking_2014,
  title = {Ranking {{Visualizations}} of {{Correlation Using Weber}}'s {{Law}}},
  author = {Harrison, Lane and Yang, Fumeng and Franconeri, Steven and Chang, Remco},
  year = 2014,
  month = dec,
  journal = {IEEE Transactions on Visualization and Computer Graphics},
  volume = {20},
  number = {12},
  pages = {1943--1952},
  issn = {1077-2626},
  doi = {10.1109/TVCG.2014.2346979},
  urldate = {2026-06-19},
  copyright = {https://ieeexplore.ieee.org/Xplorehelp/downloads/license-information/IEEE.html},
  langid = {english}
}

@article{harvey_domains_2019,
  title = {Domains of Cognition and Their Assessment},
  author = {Harvey, Philip D.},
  year = 2019,
  month = sep,
  journal = {Dialogues in Clinical Neuroscience},
  volume = {21},
  number = {3},
  pages = {227--237},
  issn = {1958-5969},
  doi = {10.31887/DCNS.2019.21.3/pharvey},
  urldate = {2026-06-19},
  langid = {english}
}

@inproceedings{heer_crowdsourcing_2010,
  title = {Crowdsourcing Graphical Perception: Using Mechanical Turk to Assess Visualization Design},
  shorttitle = {Crowdsourcing Graphical Perception},
  booktitle = {Proceedings of the {{SIGCHI Conference}} on {{Human Factors}} in {{Computing Systems}} ({{CHI}})},
  author = {Heer, Jeffrey and Bostock, Michael},
  year = 2010,
  pages = {203--212},
  publisher = {ACM},
  doi = {10.1145/1753326.1753357},
  urldate = {2012-12-04},
  isbn = {978-1-60558-929-9}
}

@article{hullman_hypothetical_2015,
  title = {Hypothetical {{Outcome Plots Outperform Error Bars}} and {{Violin Plots}} for {{Inferences}} about {{Reliability}} of {{Variable Ordering}}},
  author = {Hullman, Jessica and Resnick, Paul and Adar, Eytan},
  editor = {Papaleo, Elena},
  year = 2015,
  month = nov,
  journal = {PLOS ONE},
  volume = {10},
  number = {11},
  pages = {e0142444},
  issn = {1932-6203},
  doi = {10.1371/journal.pone.0142444},
  urldate = {2026-06-19},
  langid = {english}
}

@article{joslyn_decisions_2013,
  title = {Decisions {{With Uncertainty}}: {{The Glass Half Full}}},
  shorttitle = {Decisions {{With Uncertainty}}},
  author = {Joslyn, Susan and LeClerc, Jared},
  year = 2013,
  month = aug,
  journal = {Current Directions in Psychological Science},
  volume = {22},
  number = {4},
  pages = {308--315},
  publisher = {SAGE Publications Inc},
  issn = {0963-7214},
  doi = {10.1177/0963721413481473},
  urldate = {2026-06-19},
  langid = {english}
}

@article{joslyn_uncertainty_2012,
  title = {Uncertainty Forecasts Improve Weather-Related Decisions and Attenuate the Effects of Forecast Error.},
  author = {Joslyn, Susan L. and LeClerc, Jared E.},
  year = 2012,
  journal = {Journal of Experimental Psychology: Applied},
  volume = {18},
  number = {1},
  pages = {126--140},
  issn = {1939-2192, 1076-898X},
  doi = {10.1037/a0025185},
  urldate = {2026-06-19},
  langid = {english}
}

@inproceedings{kale_logic_2025,
  title = {Toward a {{Logic}} of {{Generalization}} about {{Visualization}} as a {{Decision Aid}}},
  booktitle = {2025 {{IEEE Visualization}} and {{Visual Analytics}} ({{VIS}})},
  author = {Kale, Alex},
  year = 2025,
  month = nov,
  pages = {1--5},
  issn = {2771-9553},
  doi = {10.1109/VIS60296.2025.00005},
  urldate = {2026-06-19}
}

@article{kale_visual_2021,
  title = {Visual {{Reasoning Strategies}} for {{Effect Size Judgments}} and {{Decisions}}},
  author = {Kale, Alex and Kay, Matthew and Hullman, Jessica},
  year = 2021,
  month = feb,
  journal = {IEEE Transactions on Visualization and Computer Graphics},
  volume = {27},
  number = {2},
  pages = {272--282},
  issn = {1941-0506},
  doi = {10.1109/TVCG.2020.3030335},
  urldate = {2026-06-19}
}

@misc{kay_tidybayes_2024,
  title = {Tidybayes: {{Tidy Data}} and {{Geoms}} for {{Bayesian Models}}},
  shorttitle = {Tidybayes},
  author = {Kay, Matthew},
  year = 2024,
  month = sep,
  doi = {10.5281/zenodo.13770114},
  urldate = {2026-07-01}
}

@article{kay_webers_2016,
  title = {Beyond {{Weber}}'s {{Law}}: {{A Second Look}} at {{Ranking Visualizations}} of {{Correlation}}},
  shorttitle = {Beyond {{Weber}}'s {{Law}}},
  author = {Kay, Matthew and Heer, Jeffrey},
  year = 2016,
  month = jan,
  journal = {IEEE Transactions on Visualization and Computer Graphics},
  volume = {22},
  number = {1},
  pages = {469--478},
  issn = {1941-0506},
  doi = {10.1109/TVCG.2015.2467671},
  urldate = {2026-06-19}
}

@inproceedings{kay_when_2016,
  title = {When (Ish) Is {{My Bus}}?: {{User-centered Visualizations}} of {{Uncertainty}} in {{Everyday}}, {{Mobile Predictive Systems}}},
  shorttitle = {When (Ish) Is {{My Bus}}?},
  booktitle = {Proceedings of the 2016 {{CHI Conference}} on {{Human Factors}} in {{Computing Systems}}},
  author = {Kay, Matthew and Kola, Tara and Hullman, Jessica R. and Munson, Sean A.},
  year = 2016,
  month = may,
  pages = {5092--5103},
  publisher = {ACM},
  address = {San Jose California USA},
  doi = {10.1145/2858036.2858558},
  urldate = {2026-06-19},
  isbn = {978-1-4503-3362-7},
  langid = {english}
}

@article{keller_matrices_2006,
  title = {Matrices or {{Node-Link Diagrams}}: {{Which Visual Representation}} Is {{Better}} for {{Visualising Connectivity Models}}?},
  shorttitle = {Matrices or {{Node-Link Diagrams}}},
  author = {Keller, Ren{\'e} and Eckert, Claudia M. and Clarkson, P. John},
  year = 2006,
  month = mar,
  journal = {Information Visualization},
  volume = {5},
  number = {1},
  pages = {62--76},
  issn = {1473-8716, 1473-8724},
  doi = {10.1057/palgrave.ivs.9500116},
  urldate = {2026-06-19},
  copyright = {https://journals.sagepub.com/page/policies/text-and-data-mining-license},
  langid = {english}
}

@article{landy_crowdsourcing_2020,
  title = {Crowdsourcing Hypothesis Tests: {{Making}} Transparent How Design Choices Shape Research Results},
  shorttitle = {Crowdsourcing Hypothesis Tests},
  author = {Landy, Justin F. and Jia, Miaolei (Liam) and Ding, Isabel L. and Viganola, Domenico and Tierney, Warren and Dreber, Anna and Johannesson, Magnus and Pfeiffer, Thomas and Ebersole, Charles R. and Gronau, Quentin F. and Ly, Alexander and {van den Bergh}, Don and Marsman, Maarten and Derks, Koen and Wagenmakers, Eric-Jan and Proctor, Andrew and Bartels, Daniel M. and Bauman, Christopher W. and Brady, William J. and Cheung, Felix and Cimpian, Andrei and Dohle, Simone and Donnellan, M. Brent and Hahn, Adam and Hall, Michael P. and {Jim{\'e}nez-Leal}, William and Johnson, David J. and Lucas, Richard E. and Monin, Beno{\^i}t and Montealegre, Andres and Mullen, Elizabeth and Pang, Jun and Ray, Jennifer and Reinero, Diego A. and Reynolds, Jesse and Sowden, Walter and Storage, Daniel and Su, Runkun and Tworek, Christina M. and Van Bavel, Jay J. and Walco, Daniel and Wills, Julian and Xu, Xiaobing and Yam, Kai Chi and Yang, Xiaoyu and Cunningham, William A. and Schweinsberg, Martin and Urwitz, Molly and {The Crowdsourcing Hypothesis Tests Collaboration} and Uhlmann, Eric L.},
  year = 2020,
  journal = {Psychological Bulletin},
  volume = {146},
  number = {5},
  pages = {451--479},
  publisher = {American Psychological Association},
  address = {US},
  issn = {1939-1455},
  doi = {10.1037/bul0000220}
}

@article{leclerc_cry_2015,
  title = {The {{Cry Wolf Effect}} and {{Weather-Related Decision Making}}},
  author = {LeClerc, Jared and Joslyn, Susan},
  year = 2015,
  journal = {Risk Analysis},
  volume = {35},
  number = {3},
  pages = {385--395},
  issn = {1539-6924},
  doi = {10.1111/risa.12336},
  urldate = {2026-06-19},
  langid = {english}
}

@article{loewenstein_risk_2001,
  title = {Risk as Feelings},
  author = {Loewenstein, George F. and Weber, Elke U. and Hsee, Christopher K. and Welch, Ned},
  year = 2001,
  journal = {Psychological Bulletin},
  volume = {127},
  number = {2},
  pages = {267--286},
  publisher = {American Psychological Association},
  address = {US},
  issn = {1939-1455},
  doi = {10.1037/0033-2909.127.2.267}
}

@article{maceachren_visual_2012,
  title = {Visual {{Semiotics}} \& {{Uncertainty Visualization}}: {{An Empirical Study}}},
  shorttitle = {Visual {{Semiotics}} \& {{Uncertainty Visualization}}},
  author = {MacEachren, Alan M. and Roth, Robert E. and O'Brien, James and Li, Bonan and Swingley, Derek and Gahegan, Mark},
  year = 2012,
  month = dec,
  journal = {IEEE Transactions on Visualization and Computer Graphics},
  volume = {18},
  number = {12},
  pages = {2496--2505},
  issn = {1941-0506},
  doi = {10.1109/TVCG.2012.279},
  urldate = {2026-06-19}
}

@inproceedings{mason_financial_2009,
  title = {Financial Incentives and the "Performance of Crowds"},
  booktitle = {Proceedings of the {{ACM SIGKDD Workshop}} on {{Human Computation}}},
  author = {Mason, Winter and Watts, Duncan J.},
  year = 2009,
  month = jun,
  series = {{{HCOMP}} '09},
  pages = {77--85},
  publisher = {Association for Computing Machinery},
  address = {New York, NY, USA},
  doi = {10.1145/1600150.1600175},
  urldate = {2026-06-19},
  isbn = {978-1-60558-672-4}
}

@book{mcelreath_statistical_2020,
  title = {Statistical {{Rethinking}}: {{A Bayesian Course}} with {{Examples}} in {{R}} and {{STAN}}},
  shorttitle = {Statistical {{Rethinking}}},
  author = {McElreath, Richard},
  year = 2020,
  month = mar,
  edition = {2},
  publisher = {{Chapman and Hall/CRC}},
  address = {New York},
  doi = {10.1201/9780429029608},
  isbn = {978-0-429-02960-8}
}

@article{nadav-greenberg_uncertainty_2009,
  title = {Uncertainty {{Forecasts Improve Decision Making Among Nonexperts}}},
  author = {{Nadav-Greenberg}, Limor and Joslyn, Susan L.},
  year = 2009,
  month = sep,
  journal = {Journal of Cognitive Engineering and Decision Making},
  volume = {3},
  number = {3},
  pages = {209--227},
  publisher = {SAGE Publications},
  issn = {1555-3434},
  doi = {10.1518/155534309X474460},
  urldate = {2026-06-19},
  langid = {english}
}

@inproceedings{nobre_evaluating_2020,
  title = {Evaluating {{Multivariate Network Visualization Techniques Using}} a {{Validated Design}} and {{Crowdsourcing Approach}}},
  booktitle = {Proceedings of the 2020 {{CHI Conference}} on {{Human Factors}} in {{Computing Systems}}},
  author = {Nobre, Carolina and Wootton, Dylan and Harrison, Lane and Lex, Alexander},
  year = 2020,
  month = apr,
  pages = {1--12},
  publisher = {ACM},
  address = {Honolulu HI USA},
  doi = {10.1145/3313831.3376381},
  urldate = {2026-06-19},
  isbn = {978-1-4503-6708-0},
  langid = {english}
}

@inproceedings{nobre_reading_2024,
  title = {Reading {{Between}} the {{Pixels}}: {{Investigating}} the {{Barriers}} to {{Visualization Literacy}}},
  shorttitle = {Reading {{Between}} the {{Pixels}}},
  booktitle = {Proceedings of the {{CHI Conference}} on {{Human Factors}} in {{Computing Systems}}},
  author = {Nobre, Carolina and Zhu, Kehang and M{\"o}rth, Eric and Pfister, Hanspeter and Beyer, Johanna},
  year = 2024,
  month = may,
  pages = {1--17},
  publisher = {ACM},
  address = {Honolulu HI USA},
  doi = {10.1145/3613904.3642760},
  urldate = {2026-06-19},
  isbn = {979-8-4007-0330-0},
  langid = {english}
}

@article{okoe_node-link_2019,
  title = {Node-{{Link}} or {{Adjacency Matrices}}: {{Old Question}}, {{New Insights}}},
  shorttitle = {Node-{{Link}} or {{Adjacency Matrices}}},
  author = {Okoe, Mershack and Jianu, Radu and Kobourov, Stephen},
  year = 2019,
  month = oct,
  journal = {IEEE Transactions on Visualization and Computer Graphics},
  volume = {25},
  number = {10},
  pages = {2940--2952},
  issn = {1941-0506},
  doi = {10.1109/TVCG.2018.2865940},
  urldate = {2026-06-19}
}

@article{oral_decoupling_2024,
  title = {Decoupling {{Judgment}} and {{Decision Making}}: {{A Tale}} of {{Two Tails}}},
  shorttitle = {Decoupling {{Judgment}} and {{Decision Making}}},
  author = {Oral, Ba{\c s}ak and Dragicevic, Pierre and Telea, Alexandru and Dimara, Evanthia},
  year = 2024,
  month = oct,
  journal = {IEEE Transactions on Visualization and Computer Graphics},
  volume = {30},
  number = {10},
  pages = {6928--6940},
  issn = {1941-0506},
  doi = {10.1109/TVCG.2023.3346640},
  urldate = {2026-06-19}
}

@article{padilla_effects_2017,
  title = {Effects of Ensemble and Summary Displays on Interpretations of Geospatial Uncertainty Data},
  author = {Padilla, Lace M. and Ruginski, Ian T. and {Creem-Regehr}, Sarah H.},
  year = 2017,
  month = dec,
  journal = {Cognitive Research: Principles and Implications},
  volume = {2},
  number = {1},
  pages = {40},
  issn = {2365-7464},
  doi = {10.1186/s41235-017-0076-1},
  urldate = {2026-06-19},
  langid = {english}
}

@article{padilla_uncertain_2021,
  title = {Uncertain {{About Uncertainty}}: {{How Qualitative Expressions}} of {{Forecaster Confidence Impact Decision-Making With Uncertainty Visualizations}}},
  shorttitle = {Uncertain {{About Uncertainty}}},
  author = {Padilla, Lace M. K. and Powell, Maia and Kay, Matthew and Hullman, Jessica},
  year = 2021,
  month = jan,
  journal = {Frontiers in Psychology},
  volume = {11},
  publisher = {Frontiers},
  issn = {1664-1078},
  doi = {10.3389/fpsyg.2020.579267},
  urldate = {2026-06-19},
  langid = {english}
}

@inproceedings{reda_graphical_2018,
  title = {Graphical {{Perception}} of {{Continuous Quantitative Maps}}: The {{Effects}} of {{Spatial Frequency}} and {{Colormap Design}}},
  shorttitle = {Graphical {{Perception}} of {{Continuous Quantitative Maps}}},
  booktitle = {Proceedings of the 2018 {{CHI Conference}} on {{Human Factors}} in {{Computing Systems}}},
  author = {Reda, Khairi and Nalawade, Pratik and {Ansah-Koi}, Kate},
  year = 2018,
  month = apr,
  series = {{{CHI}} '18},
  pages = {1--12},
  publisher = {Association for Computing Machinery},
  address = {New York, NY, USA},
  doi = {10.1145/3173574.3173846},
  urldate = {2026-06-19},
  isbn = {978-1-4503-5620-6}
}

@article{reda_rainbow_2023,
  title = {Rainbow {{Colormaps}}: {{What}} Are {{They Good}} and {{Bad}} For?},
  shorttitle = {Rainbow {{Colormaps}}},
  author = {Reda, Khairi},
  year = 2023,
  month = dec,
  journal = {IEEE Transactions on Visualization and Computer Graphics},
  volume = {29},
  number = {12},
  pages = {5496--5510},
  issn = {1941-0506},
  doi = {10.1109/TVCG.2022.3214771},
  urldate = {2026-06-19}
}

@article{reda_rainbows_2021,
  title = {Rainbows {{Revisited}}: {{Modeling Effective Colormap Design}} for {{Graphical Inference}}},
  shorttitle = {Rainbows {{Revisited}}},
  author = {Reda, Khairi and Szafir, Danielle Albers},
  year = 2021,
  month = feb,
  journal = {IEEE Transactions on Visualization and Computer Graphics},
  volume = {27},
  number = {2},
  pages = {1032--1042},
  issn = {1941-0506},
  doi = {10.1109/TVCG.2020.3030439},
  urldate = {2026-06-19}
}

@article{rensink_perception_2010,
  title = {The {{Perception}} of {{Correlation}} in {{Scatterplots}}},
  author = {Rensink, Ronald A. and Baldridge, Gideon},
  year = 2010,
  journal = {Computer Graphics Forum},
  volume = {29},
  number = {3},
  pages = {1203--1210},
  issn = {1467-8659},
  doi = {10.1111/j.1467-8659.2009.01694.x},
  urldate = {2026-06-19},
  langid = {english}
}

@article{retchless_guidance_2016,
  title = {Guidance for Representing Uncertainty on Global Temperature Change Maps},
  author = {Retchless, David P. and Brewer, Cynthia A.},
  year = 2016,
  journal = {International Journal of Climatology},
  volume = {36},
  number = {3},
  pages = {1143--1159},
  issn = {1097-0088},
  doi = {10.1002/joc.4408},
  urldate = {2026-06-19},
  langid = {english}
}

@misc{RLang_2024,
  title = {R: {{A Language}} and {{Environment}} for {{Statistical Computing}}},
  author = {R Core Team},
  year = 2024,
  address = {Vienna, Austria},
  howpublished = {R Foundation for Statistical Computing}
}

@article{rosenthal_file_1979,
  title = {The File Drawer Problem and Tolerance for Null Results},
  author = {Rosenthal, Robert},
  year = 1979,
  journal = {Psychological Bulletin},
  volume = {86},
  number = {3},
  pages = {638--641},
  publisher = {American Psychological Association},
  address = {US},
  issn = {1939-1455},
  doi = {10.1037/0033-2909.86.3.638}
}

@article{ruginski_non-expert_2016,
  title = {Non-Expert Interpretations of Hurricane Forecast Uncertainty Visualizations},
  author = {Ruginski, Ian T. and Boone, Alexander P. and Padilla, Lace M. and Liu, Le and Heydari, Nahal and Kramer, Heidi S. and Hegarty, Mary and Thompson, William B. and House, Donald H. and {Creem-Regehr}, Sarah H.},
  year = 2016,
  month = apr,
  journal = {Spatial Cognition \& Computation},
  volume = {16},
  number = {2},
  pages = {154--172},
  issn = {1387-5868, 1542-7633},
  doi = {10.1080/13875868.2015.1137577},
  urldate = {2026-06-19},
  langid = {english}
}

@inproceedings{sarma_milliways_2024,
  title = {Milliways: {{Taming Multiverses}} through {{Principled Evaluation}} of {{Data Analysis Paths}}},
  shorttitle = {Milliways},
  booktitle = {Proceedings of the {{CHI Conference}} on {{Human Factors}} in {{Computing Systems}}},
  author = {Sarma, Abhraneel and Hwang, Kyle and Hullman, Jessica and Kay, Matthew},
  year = 2024,
  month = may,
  pages = {1--15},
  publisher = {ACM},
  address = {Honolulu HI USA},
  doi = {10.1145/3613904.3642375},
  urldate = {2026-06-19},
  isbn = {979-8-4007-0330-0},
  langid = {english}
}

@inproceedings{sarma_more_2025,
  title = {More {{Forecasts}}, {{More}} ({{Decision}}) {{Problems}}: {{How Uncertainty Representations}} for {{Multiple Forecasts Impact Decision Making}}},
  shorttitle = {More {{Forecasts}}, {{More}} ({{Decision}}) {{Problems}}},
  booktitle = {Proceedings of the 2025 {{CHI Conference}} on {{Human Factors}} in {{Computing Systems}}},
  author = {Sarma, Abhraneel and Hedayati, Maryam and Kay, Matthew},
  year = 2025,
  month = apr,
  series = {{{CHI}} '25},
  pages = {1--14},
  publisher = {Association for Computing Machinery},
  address = {New York, NY, USA},
  doi = {10.1145/3706598.3713725},
  urldate = {2026-08-06},
  isbn = {979-8-4007-1394-1}
}

@inproceedings{sarma_multiverse_2023,
  title = {Multiverse: {{Multiplexing Alternative Data Analyses}} in {{R Notebooks}}},
  shorttitle = {Multiverse},
  booktitle = {Proceedings of the 2023 {{CHI Conference}} on {{Human Factors}} in {{Computing Systems}}},
  author = {Sarma, Abhraneel and Kale, Alex and Moon, Michael Jongho and Taback, Nathan and Chevalier, Fanny and Hullman, Jessica and Kay, Matthew},
  year = 2023,
  month = apr,
  pages = {1--15},
  publisher = {ACM},
  address = {Hamburg Germany},
  doi = {10.1145/3544548.3580726},
  urldate = {2026-06-19},
  isbn = {978-1-4503-9421-5},
  langid = {english}
}

@inproceedings{sarma_odds_2024,
  title = {Odds and {{Insights}}: {{Decision Quality}} in {{Exploratory Data Analysis Under Uncertainty}}},
  shorttitle = {Odds and {{Insights}}},
  booktitle = {Proceedings of the 2024 {{CHI Conference}} on {{Human Factors}} in {{Computing Systems}}},
  author = {Sarma, Abhraneel and Pu, Xiaoying and Cui, Yuan and Correll, Michael and Brown, Eli T and Kay, Matthew},
  year = 2024,
  month = may,
  series = {{{CHI}} '24},
  pages = {1--14},
  publisher = {Association for Computing Machinery},
  address = {New York, NY, USA},
  doi = {10.1145/3613904.3641995},
  urldate = {2026-06-19},
  isbn = {979-8-4007-0330-0}
}

@inproceedings{sarma_tasks_2024,
  title = {Tasks and {{Telephones}}: {{Threats}} to {{Experimental Validity}} Due to {{Misunderstandings}} of {{Visualisation Tasks}} and {{Strategies Position Paper}}},
  shorttitle = {Tasks and {{Telephones}}},
  booktitle = {2024 {{IEEE Evaluation}} and {{Beyond}} - {{Methodological Approaches}} for {{Visualization}} ({{BELIV}})},
  author = {Sarma, Abhraneel and Long, Sheng and Correll, Michael and Kay, Matthew},
  year = 2024,
  month = oct,
  pages = {33--40},
  doi = {10.1109/BELIV64461.2024.00009},
  urldate = {2026-06-19}
}

@article{simmons_false-positive_2011,
  title = {False-{{Positive Psychology}}: {{Undisclosed Flexibility}} in {{Data Collection}} and {{Analysis Allows Presenting Anything}} as {{Significant}}},
  shorttitle = {False-{{Positive Psychology}}},
  author = {Simmons, Joseph P. and Nelson, Leif D. and Simonsohn, Uri},
  year = 2011,
  month = nov,
  journal = {Psychological Science},
  volume = {22},
  number = {11},
  pages = {1359--1366},
  publisher = {SAGE Publications Inc},
  issn = {0956-7976},
  doi = {10.1177/0956797611417632},
  urldate = {2026-06-19},
  langid = {english}
}

@article{simonsohn_specification_2020,
  title = {Specification Curve Analysis},
  author = {Simonsohn, Uri and Simmons, Joseph P. and Nelson, Leif D.},
  year = 2020,
  month = nov,
  journal = {Nature Human Behaviour},
  volume = {4},
  number = {11},
  pages = {1208--1214},
  publisher = {Nature Publishing Group},
  issn = {2397-3374},
  doi = {10.1038/s41562-020-0912-z},
  urldate = {2026-06-19},
  copyright = {2020 The Author(s), under exclusive licence to Springer Nature Limited},
  langid = {english}
}

@article{steegen_increasing_2016,
  title = {Increasing {{Transparency Through}} a {{Multiverse Analysis}}},
  author = {Steegen, Sara and Tuerlinckx, Francis and Gelman, Andrew and Vanpaemel, Wolf},
  year = 2016,
  month = sep,
  journal = {Perspectives on Psychological Science},
  volume = {11},
  number = {5},
  pages = {702--712},
  publisher = {SAGE Publications Inc},
  issn = {1745-6916},
  doi = {10.1177/1745691616658637},
  urldate = {2026-06-19},
  langid = {english}
}

@article{stevens_psychophysical_1957,
  title = {On the Psychophysical Law},
  author = {Stevens, S. S.},
  year = 1957,
  journal = {Psychological Review},
  volume = {64},
  number = {3},
  pages = {153--181},
  publisher = {American Psychological Association},
  address = {US},
  issn = {1939-1471},
  doi = {10.1037/h0046162}
}

@article{szafir_modeling_2018,
  title = {Modeling {{Color Difference}} for {{Visualization Design}}},
  author = {Szafir, Danielle Albers},
  year = 2018,
  month = jan,
  journal = {IEEE Transactions on Visualization and Computer Graphics},
  volume = {24},
  number = {1},
  pages = {392--401},
  issn = {1941-0506},
  doi = {10.1109/TVCG.2017.2744359},
  urldate = {2026-06-19}
}

@article{wicherts_degrees_2016,
  title = {Degrees of {{Freedom}} in {{Planning}}, {{Running}}, {{Analyzing}}, and {{Reporting Psychological Studies}}: {{A Checklist}} to {{Avoid}} p-{{Hacking}}},
  shorttitle = {Degrees of {{Freedom}} in {{Planning}}, {{Running}}, {{Analyzing}}, and {{Reporting Psychological Studies}}},
  author = {Wicherts, Jelte M. and Veldkamp, Coosje L. S. and Augusteijn, Hilde E. M. and Bakker, Marjan and {van Aert}, Robbie C. M. and {van Assen}, Marcel A. L. M.},
  year = 2016,
  month = nov,
  journal = {Frontiers in Psychology},
  volume = {7},
  publisher = {Frontiers},
  issn = {1664-1078},
  doi = {10.3389/fpsyg.2016.01832},
  urldate = {2026-06-19},
  langid = {english}
}

@article{wu_rational_2024,
  title = {The {{Rational Agent Benchmark}} for {{Data Visualization}}},
  author = {Wu, Yifan and Guo, Ziyang and Mamakos, Michalis and Hartline, Jason and Hullman, Jessica},
  year = 2024,
  month = jan,
  journal = {IEEE Transactions on Visualization and Computer Graphics},
  volume = {30},
  number = {1},
  pages = {338--347},
  issn = {1941-0506},
  doi = {10.1109/TVCG.2023.3326513},
  urldate = {2026-06-19}
}

@inproceedings{yang_subjective_2023,
  title = {Subjective {{Probability Correction}} for {{Uncertainty Representations}}},
  booktitle = {Proceedings of the 2023 {{CHI Conference}} on {{Human Factors}} in {{Computing Systems}}},
  author = {Yang, Fumeng and Hedayati, Maryam and Kay, Matthew},
  year = 2023,
  month = apr,
  series = {{{CHI}} '23},
  pages = {1--17},
  publisher = {Association for Computing Machinery},
  address = {New York, NY, USA},
  doi = {10.1145/3544548.3580998},
  urldate = {2026-06-19},
  isbn = {978-1-4503-9421-5}
}

@article{zhang_ubiquitous_2012,
  title = {Ubiquitous {{Log Odds}}: {{A Common Representation}} of {{Probability}} and {{Frequency Distortion}} in {{Perception}}, {{Action}}, and {{Cognition}}},
  shorttitle = {Ubiquitous {{Log Odds}}},
  author = {Zhang, Hang and Maloney, Laurence T.},
  year = 2012,
  month = jan,
  journal = {Frontiers in Neuroscience},
  volume = {6},
  publisher = {Frontiers},
  issn = {1662-453X},
  doi = {10.3389/fnins.2012.00001},
  urldate = {2026-06-19},
  langid = {english}
}

\pagebreak

\appendix 

\section{Interpreting the Results of Experiment 2}
\label{appendix:exp2}

\customsubsubsection{Calculation of Benchmarks}\\
As described in \autoref{sec:exp2}, there are two payoff relevant states (of nature), $S = \{s_1 = Pr(T \leq 0), s_2 = Pr(T > 0)\}$ and two possible actions $A = \{a_1, a_2\}$. 
Following prior studies in decision-making under uncertainty in visualization~\cite[e.g.,][]{fernandes_uncertainty_2018, kale_visual_2021, padilla_uncertain_2021}, to evaluate participant decisions for each trial, we simulate a draw from the visualized temperature distribution which determines the state of the world $S$ (i.e., whether the temperature is below or above freezing). We then calculate the cost incurred based on the payoff matrix described in \autoref{sec:exp2}. To assess participants performance, we use three benchmarks---the expected utility of decision-maker responding at random, the expected utility of decision-maker with extreme risk-aversion (deciding to salt on every trial except the attention check ones), and the expected utility of completely rationally decision-maker.

\vspace{-16pt}
\begin{align*}
& \mathbb{E}(U | \text{random}) \hskip1.25em = -1000 \cdot \left( \sum_k^n 0.5 \cdot 1 + 0.5 \cdot 5 \cdot p_k \right) \\
&\hskip6.8em = -3272 \\
& \mathbb{E}(U | \text{risk-averse}) = \sum_i^n -1000 \\
&\hskip6.9em = 0 \\
& \mathbb{E}(U | \text{rational}) \hskip1.29em = -1000 \cdot \left( \sum_k^n 1 \cdot \mathbb{I}(p_i \geq 0.2) + 5 \cdot \mathbb{I}(p_i < 0.2) \cdot p_k \right) \\
&\hskip6.8em = 3960 \\
\end{align*}
\vspace{-28pt}

\noindent where $p_k$ represents the probability of freezing in a particular trial $k$ and lies in $\{0.595, 0.5, 0.405, 0.315, 0.235, 0.168, 0.115, 0.075, 0.046\}$ each repeated twice, and $n = 18$.

\vspace{4pt}
\customsubsubsection{Calculation of the Posterior Predictive Credible Interval}\\
In \autoref{fig:dist-utility}, we show the posterior predictive credible interval for expected utility in each condition based on the model estimated probability of salting for the average participant on a specific trial $k$: $p_{\textsc{salt}} = \text{logit}^{-1}(\alpha_i + \beta_i \cdot [ \text{logit}(p_k) - \text{logit}(0.2) ])$. The expected utility is given by:

\vspace{-16pt}
\[
\mathbb{E}(U) = -1000 \cdot \left( \sum_k^n \mathbb{E}(p_{\textsc{salt}}) + 5 \cdot (1 - \mathbb{E}(p_{\textsc{salt}})) \cdot p_k \right) \\
\]
\vspace{-8pt}

\noindent We sample draws from the expectation of the posterior predictive distribution using the \texttt{add\_epred\_draws} function from \texttt{tidybayes 3.0.7}~\cite{kay_tidybayes_2024}.

\vspace{4pt}
\customsubsubsection{The Probability of Performing Better than the Rational Benchmark}
In \autoref{fig:dist-utility}, we find that approximately 30\% of the participants performed better than the rational benchmark. While on the face of it, this might seem surprising, it is actually not that unlikely---the rational benchmark is an asymptotic guarantee given an infinite number of trials; for obvious purposes, we limit participants to 18 trials. To perform better than the rational benchmark ($\mathbb{E}(U) > \mathbb{E}(U | \text{rational})$), a participant simply has to get lucky and not encounter a freezing state of nature (i.e., $s_1$) on any of the trials where they decide to not salt. This is given by:

\vspace{-10pt}
\[
\prod_k^n (1 - p_i) \cdot \mathbb{I}(p_i \leq d)
\]
\vspace{-8pt}

\noindent where $d$ is the decision boundary and $d = 0.2$ for the rational decision-maker. The probability of outperforming the rational benchmark for a participant making decisions using $d = 0.2$ is 0.42. If $d = 0.3$, the probability of outperforming the rational benchmark is 0.25.\footnote{This is actually the lower bound probability. In reality, if $d \geq 0.3$, the participant can actually encounter a freezing state of nature for one of the trials and still end up with utility greater than the rational benchmark. Such a participant will salt on eight of the trials where $p_k \geq 0.3$ leaving them with a budget of \$10,000; if they encounter a freezing state of nature in one of the remaining trials, their remaining budget will be \$5,000 which is still greater than the rational benchmark}

\section{Experiment 3: Decision-Making under Uncertainty}
\label{appendix:exp3}
We conducted a replication of Experiment 2, where we examined the impact of incentives on performance in the same decision-making task using either 95\% interval (as opposed to 66\% and 95\% interval) or density plots as uncertainty representations. The rest of the experimental materials (task, procedure, tutorials and training) and data analysis model were the same as Experiment 2. As this experiment was conducted after the initial review cycle, we only include it as an appendix.

\customsubsubsection{Participants:} We recruited all participants from Prolific. Our experiment was only eligible to participants who were fluent in English, and on desktop devices. As per our pre-registrations, we aimed to recruit 240 participants (60 participants in each condition). After excluding participants who failed to meet our pre-registered attention check criteria (eleven), we had 229 participants (56 in the \baseci{}, 60 in \incci, 55 in \basedens, and 58 in \incdens). We received explicit consent from all participants to collect and share their responses. The median completion time for participants in the baseline condition was approximately 13 mins, and they were compensated \$3.5 (approximately \$16/h); the median completion time for participants in the incentivized condition was approximately 13.5 mins, and they received a guaranteed amount of \$1.9 (adjusted up from \$1.5) and the average bonus was \$1.5 (corresponding to an average wage of \$15/h).

\begin{figure}[b!]
    \vspace{-14pt}
    \centering
    \includegraphics[width=0.98\columnwidth]{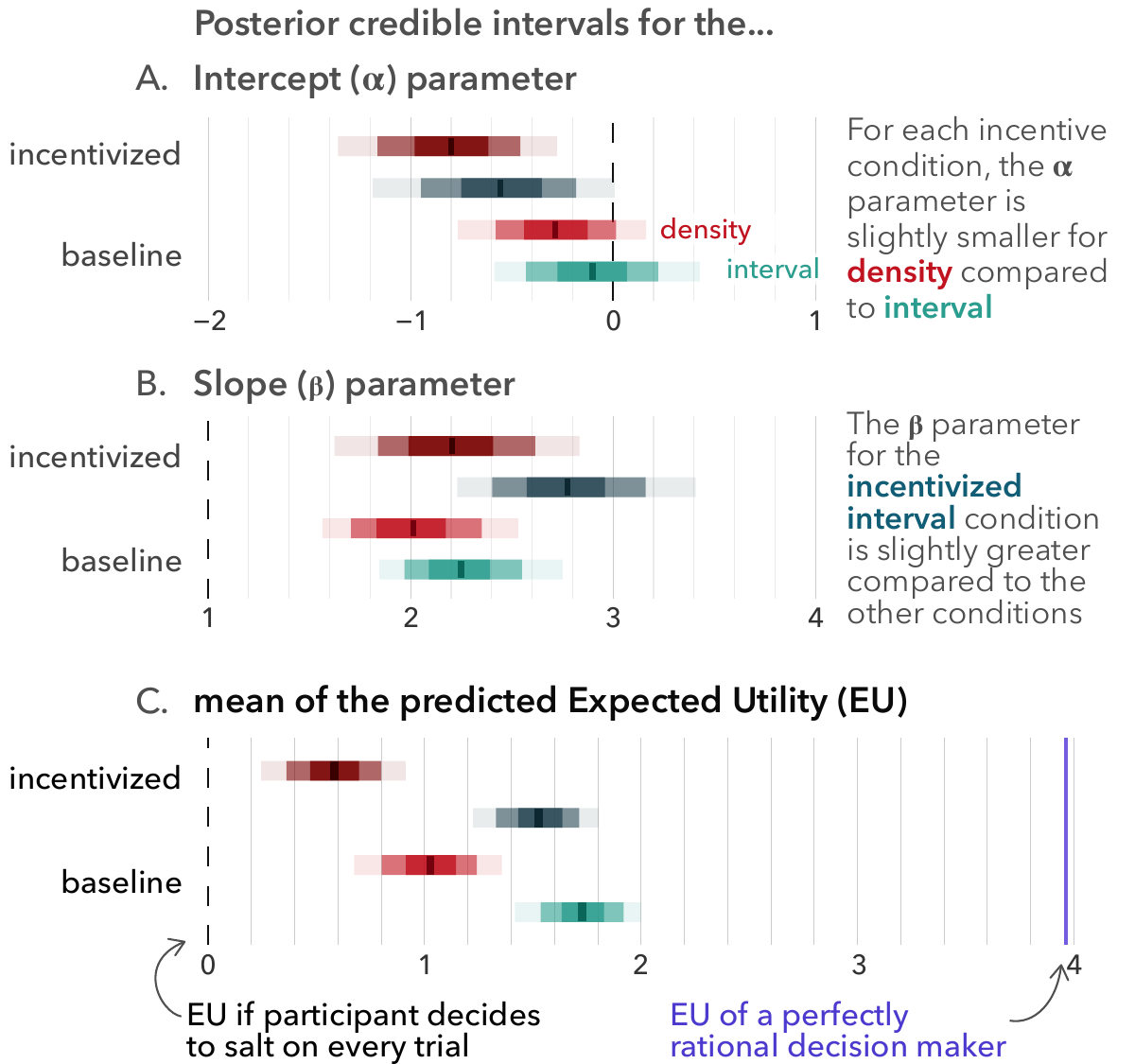}
    \vspace{-6pt}
    \caption{The main result of Experiment 3. We show the posterior credible intervals of $\alpha$, $\beta$ and the mean expected utility for both visual representations across the incentivized and baseline conditions.}
    \label{fig:dm2-results}
    \vspace{-4pt}
\end{figure}

The results from this experiment are shown in \autoref{fig:dm2-results}. Similar to Experiment 2, we found the value of $\alpha$ to be negative across all conditions (\autoref{fig:dm2-results}A)---this means that the crossover point for the average participant is greater than 0.2, suggesting a consistent bias towards risk-seeking behavior. We also found the value of $\beta$ to be significantly larger than one across all conditions (\autoref{fig:dm2-results}B), and highest for the \incci{} condition, indicating that participants are quite sensitive to the stimuli and their subjective crossover point. \autoref{fig:dm2-results}C allows to compare decision quality. We again do not find any meaningful effect of incentives. In addition, we found the expected utility to be greater for both interval conditions (\baseci{: 1.73, [1.42, 2.00]} and \incci{: 1.53, [1.22, 1.80]}) compared to the density conditions (\basedens{: 1.03, [0.67, 1.36]} and \incdens{: 0.58, [0.24, 0.91]}). While this result is consistent with our previous experiment, it is contrary to the findings of prior work which have generally found that visualizing uncertainty using density plots lead to better decision quality.

As in the previous two studies, we again find that participants took slightly longer to complete the task in the incentivized conditions compared to the baseline conditions. \autoref{fig:time-taken}B, visualizes the bootstrapped median and 95\% quantile intervals for the median estimate. The difference in time spent is approximately 22s for intervals (\baseci{: 179s, [160s, 201s]}, \incci{: 201s, [170s, 217s]}), and approximately 30s for density  (\basedens{: 194s, [161s, 239s]}, \incdens{: 224s, [186s, 292s]}). This increase of 10-15\% is smaller than what was observed for Experiment 2, and similar to the difference observed in Experiment 1.






\end{document}